\documentclass[12pt,preprint]{aastex}
\bibpunct{(}{)}{;}{a}{}{,} 
\usepackage[section] {placeins}
\usepackage{graphicx}
\usepackage{epstopdf}
\shorttitle{Proper Motion Study of the Magellanic Clouds using SPM material}
\shortauthors{Vieira, et al.}

\begin{document}

\title{Proper Motion Study of the Magellanic Clouds using SPM material}

\author{Katherine Vieira\footnote{Now at Centro de Investigaciones de Astronomia,
Apartado Postal 264, M\'erida 5101-A, Venezuela}, Terrence M. Girard, William F. van Altena}
\affil{Astronomy Department, Yale University, P.O. Box 208101, New Haven, CT 06520, USA}
\email{kvieira@cida.ve,terry.girard,william.vanaltena@yale.edu}

\author{Norbert Zacharias}
\affil{US Naval Observatory, 3450 Massachusetts Av. NW, Washington, DC 20392, USA}
\email{nz@usno.navy.mil}

\author{Dana I. Casetti-Dinescu, Vladimir I. Korchagin}
\affil{Astronomy Department, Yale University, P.O. Box 208101, New Haven, CT 06520, USA}
\email{dana.casetti@yale.edu,vkorchagin@sfedu.ru}

\author{Imants Platais}
\affil{Department of Physics and Astronomy, Johns Hopkins University, 3400 North Charles Street, Baltimore, MD 21218, USA}
\email{imants@pha.jhu.edu}

\author{David G. Monet}
\affil{US Naval Observatory, Flagstaff Station, P.O. Box 1149, Flagstaff, AZ 86002}
\email{dgm@nofs.navy.mil}

\author{Carlos E. L\'opez}
\affil{Universidad de San Juan and Yale Southern Observatory, Avenida Benavidez 8175 Oeste, Chimbas, 5413 San Juan, Argentina}
\email{cel$\_$2018@yahoo.com.ar}

\begin{abstract}

Absolute proper motions are determined for stars and galaxies to V=17.5
over a 450 square-degree area that encloses both Magellanic Clouds.
The proper motions are based on photographic and CCD observations of the
Yale/San Juan Southern Proper Motion program, which span over a baseline of 40 years.
Multiple, local relative proper motion measures are combined in an overlap
solution using photometrically selected Galactic Disk stars to define
a global relative system that is then transformed to absolute using external
galaxies and Hipparcos stars to tie into the ICRS.
The resulting catalog of 1.4 million objects is used to derive the
mean absolute proper motions of the Large Magellanic Cloud and the
Small Magellanic Cloud; 
$(\mu_\alpha\cos\delta,\mu_\delta)_{LMC}=(1.89,+0.39)\pm (0.27,0.27)\;\;\mbox{mas yr}^{-1}$ and
$(\mu_\alpha\cos\delta,\mu_\delta)_{SMC}=(0.98,-1.01)\pm (0.30,0.29)\;\;\mbox{mas yr}^{-1}$.
These mean motions are based on best-measured samples of 3822 LMC stars and
964 SMC stars.
A dominant portion (0.25 mas~yr$^{-1}$) of the formal errors is due to the 
estimated uncertainty
in the inertial system of the Hipparcos Catalog stars used to anchor the bright
end of our proper motion measures.
A more precise determination can be made for the proper motion of the SMC
{\it relative} to the LMC;
$(\mu_{\alpha\cos\delta},\mu_\delta)_{SMC-LMC} = (-0.91,-1.49) \pm (0.16,0.15)\;\;\mbox{mas yr}^{-1}$.
This differential value is combined with measurements of the proper motion 
of the LMC taken from the literature to produce new absolute
proper-motion determinations for the SMC, as well as an estimate of the
total velocity difference of the two clouds to within $\pm$54 kms$^{-1}$.
The absolute proper motion results are consistent with the Clouds' orbits 
being marginally bound to the Milky Way, albeit on an elongated orbit.
The inferred relative velocity between the Clouds places them near their
binding energy limit and, thus, no definitive conclusion can be made
as to whether or not the Clouds are bound to one another.

\end{abstract}


\section{Introduction}\label{sc_intro}

The Magellanic Clouds have provided astronomers with a wide variety of information,
from the small to the large hierarchy of objects in the Universe.
They are the first step in the cosmic-distance ladder,
as well as a proxy for the low-metal
gas-rich galaxies assembled in the early universe, they have zones of recent massive star 
formation, young 1-3 Gyr globular clusters and a variety of old pulsating stars and planetary nebulae.
They contain substantial amounts of gas and dust in a violent interstellar medium, 
harbor the closest supernova in recent years, and are being used as a testbed
for dark matter searches.

The Magellanic Clouds are also the prime example of a galaxy-galaxy interaction, 
based on several lines of evidence, close enough to be studied in detail:  
An apparently starless Magellanic Stream trails the Clouds \citep{1974ApJ...190..291M,1982Obs...102..174P};  
a bridge of gas and stars connects the Clouds \citep{1963AuJPh..16..570H,1985Natur.318..160I}; 
a gaseous Leading Arm precedes the clouds \citep{1998Natur.394..752P};
an HI envelope surrounds the whole system and 
a collection of High Velocity Clouds seems to be ``raining'' over the Galaxy \citep{2004A&A...423..895O}. 
Despite such convenient observational circumstances, on the theoretical side
no dynamical model has been able to reproduce all these phenomena simultaneously.

It is widely accepted that the complexity and intricacies of the Magellanic Clouds'
external and internal features, have been largely determined by the orbit they have
followed in the past few Gigayears. Due to their large distance, about 50 and 60 kpc to
the LMC and the SMC, respectively, only radial velocities 
were precise enough to provide some assessment of their spatial
velocity. In fact, line-of-sight measurements of the Clouds began
about a hundred years ago \citep{1915PNAS....1..183W}, 
while proper motion measurements of a useful accuracy were possible only in the 1990's.

The first proper motion results 
\citep{1989BAAS...21.1107J,1991IAUS..148..491T,1993AGAb....8..155B,
1993BAAS...25..783L,1994MNRAS.266..412K,1994AJ....107.1333J,
1996BAAS...28..932I,1997NewA....2...77K,1997AGAb...13...77K,
1997ESASP.402..615K,1999IAUS..190..475A} based on 
plate and/or CCD data, were compatible with a picture in which
the Magellanic Clouds were bound to each other and to the
Milky Way. Such scenario relied heavily on the fact that the
Galactic gravitational potential used (isothermal sphere) 
yields such results by default, and proper motion errors were not
small enough to refine the tangential velocities.

In the past ten years though, investigations yielded quite a variety of
results \citep{2000AJ....120..845A,2001AAS...199.5205D,2002AJ....123.1971P,
2005A&A...437..339M,2005AAS...20711307K,2006ApJ...638..772K,
2006AJ....131.1461P,2006RMxAC..25...43P,2006RMxAC..26...78P,2006RMxAC..26..183M,
2006ApJ...652.1213K,2008AJ....135.1024P,2009IAUS..256...93K,2009AJ....137.4339C}.
Some \citep{2000AJ....120..845A,2005A&A...437..339M} were found to have 
unknown or important systematic errors. 
More interestingly, HST-based results \citep{2005AAS...20711307K,2006ApJ...638..772K,
2006ApJ...652.1213K,2008AJ....135.1024P,2009IAUS..256...93K} coupled to more
modern and cosmologically inspired dark matter Halo models, suggest
that the Clouds are not bound to the Galaxy, opposite to the long-held paradigm.

Twenty years have passed since the very first proper motion measurements of the Clouds, 
and it is only now with the Yale-San Juan Southern Proper Motion 
(SPM) program - briefly explained in Section \ref{sc_spm} - that for the 
first time a wide-field astrometric 
proper motion survey of the Magellanic System is finally completed.
All the 1st-epoch (early 1970's) and part of the 2nd-epoch (early 1990's)
SPM material used in this work are photographic plates.
Their processing is briefly summarised in Section \ref{sc_plate}, but a more
detailed explanation can be found in Girard et al. (2010).

A substantial part of our 2nd-epoch data comes from SPM CCD observations.
A short explanation of the data acquisition, 
quality control and processing is explained in Section
\ref{sc_ccd}. To supplement our 2nd-epoch plate data with CCD-quality
positions, we have included mean positions at Julian epoch 2000.0 
from the UCAC2 catalog \citep{2004AJ....127.3043Z}.
Section \ref{sc_ucac2} explains how these data were selected
and included in this investigation.

In contrast to other SPM reductions, as explained in Section \ref{sc_obtpm},
relative proper motions measured in CCD-size fields of view were combined
into a single common extended and accurately defined global system.
Although our zero point accuracy, i.e. how well our reference frame is linked 
to the International Celestial Reference System (ICRS), is ultimately 
limited by Hipparcos accuracy, our very precise relative proper motions 
over the whole field of view, enabled us to measure the proper motion of the SMC 
with respect to the LMC, at a precision comparable to the quoted errors of 
space-based proper motions. 

It is this capability that we exploit to obtain new measurements of the proper motion
of the SMC based on previously published LMC proper motions. Section \ref{sc_pmmcs} contains
the main results of this paper regarding the proper motion of the Clouds, absolute and relative.
Section \ref{sc_implications} has a discussion of the implications of our results 
on the current understanding of the dynamics of the Magellanic System.
Finally, Section \ref{sc_end} states the conclusions of this paper, and 
future plans already in consideration to improve the current results.

\section{The SPM program}\label{sc_spm}

This investigation is part of the SPM program, a joint venture of Yale University
and Universidad Nacional de San Juan in Argentina. 
The SPM program was initiated in the
early 1960s by D. Brower and J. Schilt as a joint enterprise of
the Yale and Columbia Universities \citep{1974IAUS...61..201W}.
The goal of the SPM program is to provide absolute proper motions, positions, 
and $BV$ photometry for the Southern sky to a limiting magnitude of $V\sim 18$.

The SPM program makes use of the Yale Southern Observatory's double astrograph 
at Cesco Observatory in the foothills of the Andes mountains in El Leoncito,
Argentina. This telescope consists of two 51-cm refractors, 
designed for photography in the blue and yellow bands, respectively.
The first-epoch survey, taken between 1965 and 1974 was made on 
glass photographic plates, exposed simultaneously in blue-yellow pairs 
and always centered on the meridian. The plates' field of view (FOV) 
extends over an area of $6.3^o \times 6.3^o$. 
The sky south of $\delta=-17^o$ was observed in the first-epoch period.

Second-epoch SPM plate observations were begun in 1988. By the mid 
1990's, with only a third of the second-epoch survey completed,
Kodak discontinued the production of the photographic plates. 
In 2000, with funding from the NSF, a CCD camera system was installed on the 
double astrograph to replace
the photographic plate holders. A PixelVision 4K$\times$4K CCD camera
($0.94^o\times 0.94^o$ FOV) was placed in the yellow
lens focal plane, and an Apogee Ap-8 1K$\times$1K ($0.37^o\times 0.37^o$ FOV) 
was fitted in the blue focal plane. In 2004, the Apogee 1K camera was
upgraded to an Apogee Alta E42 2K$\times$2K ($0.42^o\times 0.42^o$ FOV)
with funds from the Argentine CONICET.
In order to achieve a limiting magnitude similar to the first-epoch
plate material, the CCD survey consists of 2-minute exposures
in both cameras. 

In the past decade, catalogs of the SPM program covering various parts of the sky
have been published, as the second-epoch material became
available for its astrometric reduction. Catalogs SPM1 \citep{1998AJ....116.2556P}, 
SPM2 \citep{1999DDA....31.1004V}, and SPM3 \citep{2004AJ....127.3060G} 
are based on photographic plates only, in both first and
second epochs. The plates were scanned either with the Yale PDS (using
input lists of selected objects) or the USNO Precision Measuring Machine (PMM)
for the whole plate. Also, two different
centering algorithms have been used to measure the image centers in the scans,
the Yale 2D-Gaussian fit from \cite{1983AJ.....88.1683L}, or the USNO circular
fit from \cite{2003AJ....125..984M}. 

Since 2004, regular CCD observations have been carried out to finish the 
second-epoch survey of the SPM program. By December 2008,
the survey was effectively completed for the sky south of $\delta=-20^o$. 
Subsequently, the SPM4 catalog, based on all available plate
and CCD data, was completed in late 2009 and is currently available,
(Girard et al.~2010).
SPM4 includes $\sim$100 million objects south of $\delta=-20^o$,
brighter than $V\sim 18$. Many of the data-processing procedures,
software and protocols developed in this investigation were also used in the
construction of the SPM4 catalog.

\section{The Plate Data}\label{sc_plate}

\subsection{Observations}\label{ss_plateobs}

The SPM survey fields are on $5^o$ centers in declination and a maximum
of $5^o$ separation in right ascension, providing at least a full degree of
overlap between adjacent plates.
Each 17-inch $\times$ 17-inch plate covers an area of $6.3^o \times 6.3^o$ 
(55.1 "/mm plate scale) and consists of a
2-hour and a 2-minute offset exposure.
All observations were made with a wire grating over the objectives,
producing measurable diffraction images out to third order.
The grating constant is 3.8 magnitudes, thus, along with the offset
short exposure, effectively increasing the dynamic range of each plate
allowing measurement of external galaxies and bright Hipparcos stars.
Fields were observed simultaneously in blue and yellow passbands, and there
are some fields with repeated blue and/or yellow plates from the same epoch.
See \cite{2004AJ....127.3060G} for a more complete description of the SPM
plate material.

Table \ref{tab_plateslist} lists the SPM plates used in this investigation.
During the course of this research, it was found that some plates yielded unusually
deviant results. A visual examination of the suspect plates revealed that the stellar 
images suffered from significant defects, possibly caused by poor guiding, 
polar misalignment, or poor focus. 
These plates, 0750B, 0751B, 1371B, 1357Y and 1373Y,
were therefore discarded for the research presented here. Coverage
in these areas was not affected, since only one of the two plates at a given epoch 
per field affected was discarded. 
Figure \ref{fig_plates} shows the distribution on the sky of the 
plates used in this work. Twenty two (22) SPM regions were studied, of which 
seven (7) have 2nd-epoch plates. For the areas with 2nd-epoch plate data, no 2nd-epoch
CCD observations were made.

\subsection{Astrometric Reduction and Photometric Calibration}\label{plate_red}

The SPM plates used in this investigation were scanned with the Precision Measuring Machine
(PMM) of the USNO Flagstaff Station. For more details about the PMM setup and operations see 
\cite{2003AJ....125..984M}. In a collaboration between the USNO-Washington 
and the Yale Astrometry Group, the StarScan reduction pipeline \citep{2008PASP..120..644Z} was modified
to analyze the PMM pixel data of the SPM plates to  produce a list of detections,
image centers and photometric indices. 
The astrometric and photometric reductions then proceeded as follows: 
1) cross-identification of detections to an input catalog, including 
2) recognition and identification of central and higher grating orders;
3) photometric calibration to obtain $BV$; 
4) correction for Atmospheric Refraction; 
5) correction for Magnitude Equation, which also combines grating-order systems; 
6) transformation of short-exposure positions into the long-exposure system; and 
7) astrometric solution into Tycho-2 to obtain $(\alpha,\delta)$. 

The input catalog referenced in step 1) is a compilation of the following
external catalogs: 
Hipparcos \citep{1997ESASP1200.....P}, 
Tycho-2 \citep{2000A&A...355L..27H}, 
UCAC2 \citep{2004AJ....127.3043Z}, 
2MASS point-source and extended-source \citep{2006AJ....131.1163S},
LEDA galaxies with DENIS measurements \citep{2005A&A...430..751P},
and the QSO catalog of \cite{2006A&A...455..773V}.
In order for an object to be included in our study, it must appear in one
or more of these listed catalogs.
A thorough explanation of all these procedures can be found in Girard et al. (2010).
After the above processing, one has positions $(x,y)$ properly calibrated into
a common system within each plate, with computed positional errors and 
astronomical coordinates $(\alpha,\delta)$ on the ICRS, as realized by Tycho-2.

\subsection{Evaluation of the plate data}\label{ss_evaplate}

Well measured stars on the plates in all of the orders have positional errors
between 0.9 $\mu$m and 1.6 $\mu$m (50 mas and 90 mas). 
These errors, which only assess measurement uncertainties, are consistent
with the precision expected for a good centering procedure, based on
previous experience with the SPM plates.

A single final position $(x,y)$ per star per plate is obtained, from the positional-error-weighted-average 
of the available measurements. As expected, the error of the final position and 
magnitude will depend on the number of grating orders contributing to their calculation. 
If average measurements from different plates were later averaged to obtain
a final number, then other errors, random and systematic, would come into play,
and the error budget of this final result should include the additional sources
of uncertainty.

\begin{figure}
\centering
\includegraphics[width=4in,angle=270]{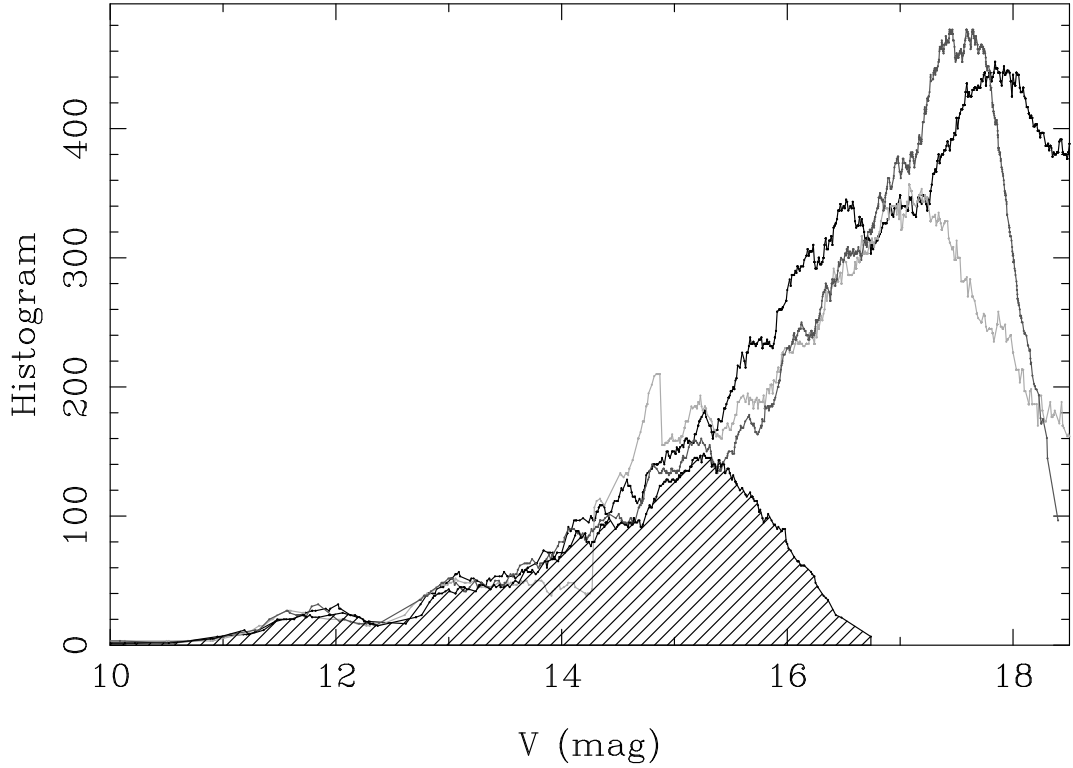}
\caption{Magnitude distribution and, thus, completeness of the SPM material evaluated in
a CCD-size field in SPM field 028 that has SPM plate data (dark grey), 
SPM CCD data (light grey) and UCAC2 data (stitch under curve),
in comparison with the 2MASS detections (black) in the field.}
\label{fig_completeness}
\end{figure}

As with earlier SPM catalogs, an 
approximate estimate of the relative completeness between the plate data and 2MASS,
can be made. Figure \ref{fig_completeness} shows the $V$ magnitude distribution of all stars
detected on a yellow plate, compared to the $V_{JK}$ magnitude distribution of all
2MASS point source stars in the same field. $V_{JK}=J+2.79(J-K)$ is an approximate empirical
transformation from 2MASS JHK to V determined by \cite{2004AJ....127.3060G}
and found to be valid for a relatively wide range of spectral types.
It can be seen that the SPM plates have a completeness similar to that
of 2MASS up to $V=17.5$ and a falloff after that.

In general, compared to previous SPM processing, these plate data have 
significantly fewer false detections, and a better correction of systematics 
in the detected positions associated with the scanning process.

\subsection{SPM 2nd-epoch plate data}\label{ss_platelet}

As explained in subsection \ref{ss_plateobs}, seven SPM fields have 2nd-epoch
plates. For this reason, no 2nd-epoch CCD observations were made at these locations. 
In order to facilitate the managing of files and software, the 2nd-epoch plates
were divided in CCD-size fields, to emulate the overlap scheme of the
2nd-epoch CCD frames (See Section \ref{ccd_obs}). This way, all the 2nd-epoch material, regardless
of its type, would have a uniform format and structure, for programming purposes.
From the about 90 CCD pointings that usually cover one SPM field, half were
used to divide the blue plate and the other half to divide the yellow
plate, in such way that these yellow and blue fields overlap in a 
similar way as do the real CCD frames.

\section{The CCD data}\label{sc_ccd}

\subsection{CCD cameras}\label{ccd_cam} 

The CCD system consists of four cameras, one each for the blue- and yellow-optimized
lenses and two focus-sensor cameras, again, one for each lens.
The CCD camera for the Yellow telescope is a PixelVision camera with a Cryotiger 
Chiller cooled (-85$^o$C) 4K $\times$ 4K Loral chip with 15 $\mu m$ pixels, which translates 
into a pixel size of 0.''83 and a total area of $0^o.94 \times 0^o.94$ degrees.
The pixel size is well matched to the YSO site where the seeing conditions usually 
yield an image FWHM of 2-3'', corresponding to 3-4 pixels per FWHM. 
This is optimal sampling for the derivation of astrometric 
image centers, based on SPM experience with digital image centering. 
The unthinned and front-illuminated Loral chip is fitted with a fixed
Custom Scientific V-band filter. 

The Blue telescope was first fitted with an Apogee AP-8 camera that utilized a 
ther\-mo\-elec\-tri\-ca\-lly-cooled (-40$^o$C) and back-illuminated 1K $\times$ 1K Site chip 
with 24 $\mu m$ pixels, which translated into a pixel size of 1.''32 and a total
area of 22.'57 $\times$ 22.'57. In May 2005, an upgrade was 
made by replacing this camera with a new Apogee Alta E42 back-illuminated 2K $\times$ 2K chip, 
with 13.5 $\mu m$ pixels, which correspond to a pixel size of 0.''74  and
a total area of 24.'8 $\times$  24.'8. 
Centered on the same field as the larger PV yellow camera, the purpose of 
the blue CCD camera is to provide B-band CCD photometry for the stars that
fall into its FOV.

Data from the yellow PV camera were used for both astrometry and photometry, 
while the blue Apogee and Alta cameras observations were used only 
for the photometric calibration of the blue plates. 

\subsection{Observations}\label{ccd_obs} 

\begin{figure}
\centering
\includegraphics[scale=0.6,angle=-90]{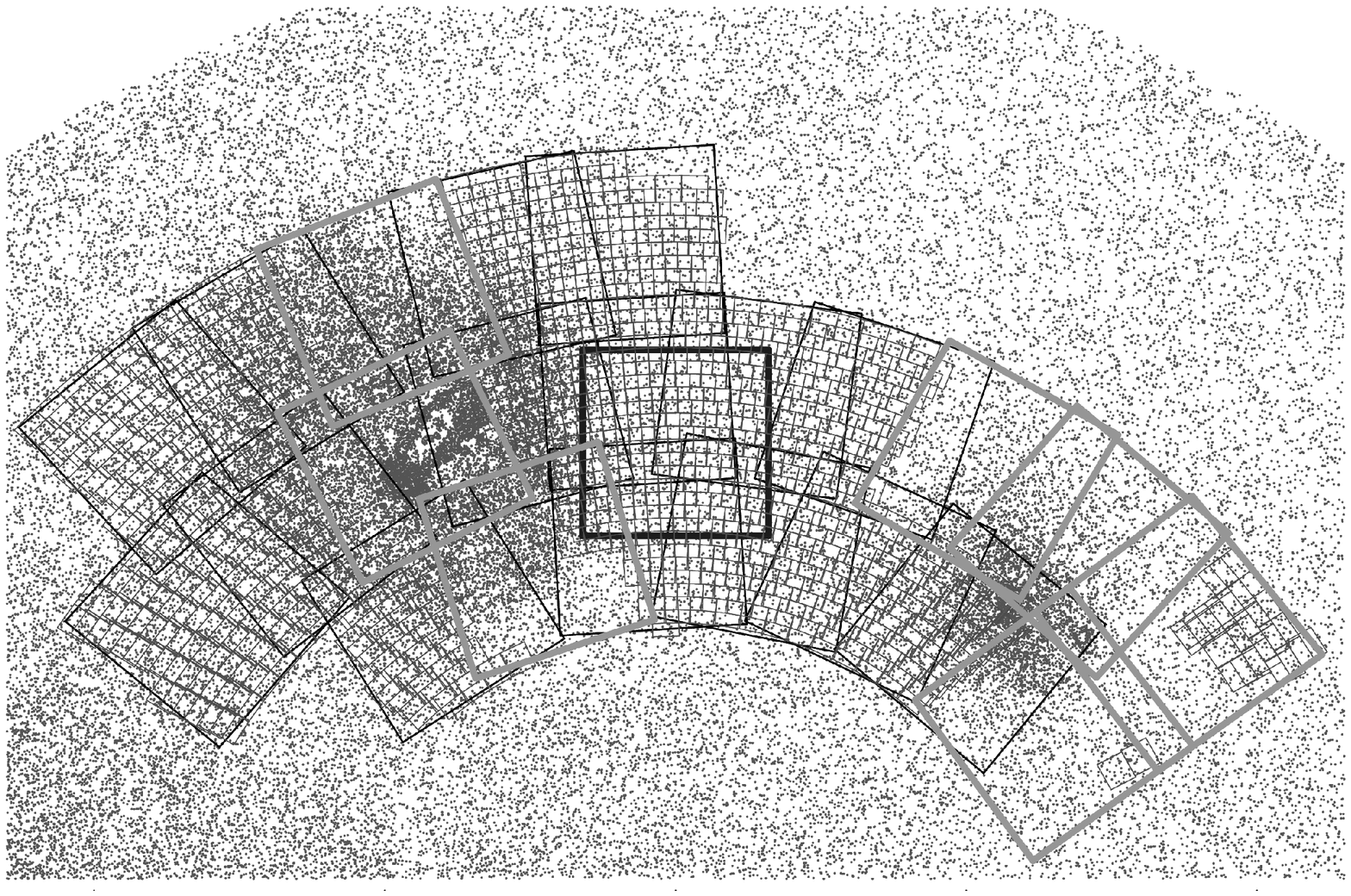}
\caption{SPM fields in the Magellanic Cloud area.
Gray outlined regions correspond to 2$^{nd}$-epoch SPM plate observations.
The small squares indicate 2$^{nd}$-epoch CCD observations.
The black outlined area is the VMC region (See Sec. \ref{ccd_obs}).
A subset of stars from the USNO-A2.0 Catalog is
plotted in the background, to indicate the positions of the 
LMC (on the left) and SMC (on the right).
SPM fields are on nominal 5$^o$ centers, not including the VMC field.}
\label{fig_plates}
\end{figure}

As a norm for the SPM, CCD observations are done always within $1^h30^m$ 
of the meridian, in 2-minute exposures, 
with the wire grating placed so that the diffraction pattern is at about $45^o$
from the E-W line. Normally, an E-W orientation is ideal to avoid differential
color refraction effects within the diffraction pattern,
but in this case, a diagonal orientation prevents the saturated central-order
image of a bright star from spoiling its grating images by either row or
column bleeding. The CCD pointings conform to
a two-fold overlap coverage scheme for the PV camera, as shown in Figure \ref{fig_plates}. 
About 90 PV CCD frames cover one single SPM field. 
Only targets that did not have 2nd-epoch plate data were 
observed with the CCD, except for a few special targets.

A total of 1310 CCD pointings were observed for this investigation. Each pointing
was planned to be observed only once, except for 90 of them
extending over a $6^o \times 6^o$ field around $(\alpha,\delta)=(3^h44^m33^s,-71^o40'18")$,
within the area delimited by the bold black line in Figure \ref{fig_plates}. 
This area corresponds to the non-SPM VMC field, 
for the Variable Stars in the Magellanic Clouds
study done from 1965 to 1968 by A. J. Wesselink,
and comprises seventy 60-minute exposure blue plates without the wire objective 
grating, reaching a limiting magnitude of about 18.
Numerous repeated CCD observations were performed on this area
to provide suitable 2nd-epoch observations for this material.
The VMC plates were not used for this work because they require their own special
reduction, different from the one used in normal SPM plates, but the repeated CCD
observations were indeed used for this investigation. The VMC CCD targets have
at least 13 good observations each, with some of them having up to 20 good observations.

The criteria used to qualify an observed frame as acceptable are: $FWHM \leq$ 3."5 , limiting magnitude
$V_{lim}\geq 17$ and standard error $\leq$ 120 mas from an astrometric solution into the UCAC2. 
If a frame fails any of these limits then it was taken again until it passed all of them.
Nonetheless, all frames regardless of quality are saved and processed,
and only later in the astrometric reduction are discarded, if they prove
to be too bad for any use. In total, 5422 CCD frames were processed 
for this investigation.

\subsection{Astrometric Reduction and Photometric Calibration}\label{ccd_ast}

Data from all the CCD cameras went through the usual processing to calibrate
the flux detected by the electronics, for the zero charge of the chip
(bias), accumulated signal from the electronics dark current (dark) and 
different response to light from each pixel (flat). Details of these
procedures can be found in Girard et al. (2010).

Image detection on the processed CCD frames is done using SExtractor 
Version 2.4.4 \citep{1996A&AS..117..393B}, from which a preliminary centroid 
and an aperture instrumental magnitude are read. 
SExtractor centroids are then used as input positions 
to compute more precise centers, based on the Yale 2D gaussian centering 
algorithm \citep{1983AJ.....88.1683L}. Typically more than 90\% 
of SExtractor detections are centered. 
A significant reduction in the rms of the positions
for repeated observations of the PV frames was found when using the
Yale-based centers as compared to the SExtractor-based centroids.

The process of transforming the Yale-based centers $(x,y)$ and 
SExtractor-based instrumental magnitudes $m_{inst}$ into calibrated 
$(\alpha,\delta)$ and $BV$, for the CCD frames, follows a procedure similar
to that of the plates (Section \ref{plate_red}), with the
exception that the external catalog used for the astrometric reduction
of the CCDs is UCAC2 (Girard et al. 2010).
As a result of these procedures, all the detected positions $(x,y)$ are properly 
calibrated within each CCD frame, with computed positional errors and equatorial 
coordinates $(\alpha,\delta)$ in the ICRS, as realized by UCAC2.

\subsection{Evaluation of the CCD data}\label{ss_evaccd}

The single-image centering precision for well measured stars ($V\leq 15$) in the CCD frames
is 0.5 $\mu$m (25 $mas$), worsening for the faintest 
stars where it reaches about 2 $\mu$m (100 $mas$). A single final position $(x,y)$ and magnitude 
per star per CCD is obtained, from the positional-error-weighted-average 
of the available measurements. The final positional and photometric errors 
will depend on the various image orders contributing to the final value.

At this point, each star in each CCD frame has a 
master ID identification, a position $(x,y)$ with errors, a calibrated 
$B$ or $V$ magnitude and a UCAC2-based $(\alpha,\delta)$.
Because stars were identified in the CCD frames in the same way as 
in the plates, similar charateristics regarding completeness
were expected when compared to 2MASS (See Figure \ref{fig_completeness}).
The CCD data show in general a completeness magnitude of about $V=18$.

\section{UCAC2 CCD-positions as supplement for the 2nd-epoch plate data}\label{sc_ucac2}

As seen in Figure \ref{fig_plates}, some fields on and adjacent to the Magellanic
Clouds have only photographic plates as 2nd-epoch material.
Given the lower quality of plate images compared to CCD 
observations, the measured proper motions in these 
plate-only fields, will have significantly larger errors than those 
coming from the combination of plate and CCD data.

In an attempt to counter this and achieve a more homogeneous quality in the final measurements,
we decided to supplement the 2nd-epoch plate measures with epoch 2000 positions
from the UCAC2 Catalog \citep{2004AJ....127.3043Z}.
These are mostly based on CCD observations\footnote{Strictly speaking, 
the supplemental UCAC2 epoch-2000 positions we use include up to a few years'
worth of UCAC2 proper motions. For the areas of sky in this study,
the UCAC2 CCD observations were made around 1998. Thus the positions 
we employ as supplements are practically those of the UCAC CCD program.} 
with the USNO 8 inch (0.2 m) Twin Astrograph from Cerro Tololo International Observatory in Chile.
The UCAC2 data were collected in such way that it mimics our 2nd-epoch fields.
The precision of the positions are 15 to 70 mas, depending on magnitude, 
with claimed estimated systematic errors of 10 mas or below. 
UCAC2 provides only crude magnitudes in a single nonstandard bandpass 
$R_{UCAC2}$ between $V$ and $R$, and its limiting magnitude is about $R\approx 16$. 

A significant number of faint stars in our 2nd-epoch SPM plates do not have a counterpart
in the UCAC2 catalog. This means we are only sampling stars about $V<16$ for these
SPM fields when using UCAC2 positions. UCAC2 completeness compared to 2MASS and SPM
data, can be seen in Figure \ref{fig_completeness}.

\section{Obtaining the proper motions}\label{sc_obtpm}

Given that a substantial part of the 2nd-epoch SPM material used in this work
comes from CCD frames, previous SPM procedures for obtaining proper motions 
(used on 1st- and 2nd-epoch plates), could not be straightforwardly applied.
Since the CCD's FOV is about 40 times smaller than the plate's FOV, the number of reference 
stars available to measure relative proper motions in each field is proportionally
smaller. A simple cut in magnitude, as in past SPM
reductions, would result in too few reference objects per CCD frame;
in particular, the extragalactic objects needed to transform relative proper motions 
into absolute ones.

However, if we select reference stars belonging to some specific population of the Galaxy, 
it is reasonable to assume that they have a mean absolute motion along 
an extended area on the sky\footnote{Stars
behave like a collisionless system, all moving under the influence of
the same general background gravitational field, therefore we can expect
them to have a global smooth distribution in their velocities, with some scatter around
a mean value at any given location.} that can be parametrized as a smooth function
of $(\alpha,\delta)$. Moreover, within a CCD FOV, their mean motion has a 
very small gradient, if any. It then becomes a matter of precisely quantifying, 
over the whole field of view, the measured mean relative proper motion of all known 
extragalactic objects, which is simply the reflex of the mean absolute proper motion 
of the reference stars. Applying such a function to the
measured relative proper motions converts them to absolute.

The layout of the CCD fields, where substantial overlap exists (20\% to 50\%), 
contains a wealth of information that can provide linkage
of the reference system across the observed region of the Clouds
that is limited only by measurement errors. A key point of this investigation 
is therefore to find a procedure to utilize this large overlap to produce a
precise global reference system.

\subsection{Relative proper motions in the small CCD FOVs}\label{sc_relpm}

After all the plate and CCD data processing described above,
the following data are ready for the measurement of proper motions:

\begin{itemize}
\item 1st-epoch positions on the plates, which have been corrected for systematics,
as much as Tycho-2 precision and number of stars available allows. 
\item 2nd-epoch positions on SPM CCD frames, SPM plates, or from UCAC2 .
CCD positions are mostly free of systematic errors. Plate positions have been corrected 
for systematics, as much as Tycho-2 stars precision and number of stars allows.
\end{itemize}

Henceforth, we will refer to each of the 2nd-epoch CCD-size frames as a {\it brick}, 
regardless of the source of its data (SPM plate, SPM CCD or UCAC2).
For each 2nd-epoch brick, the corresponding 1st-epoch plate area was reprojected
onto a tangential plane centered on the brick. Then a quadratic solution was computed to
transform the reprojected 1st-epoch plate's $(x,y)$ into the 2nd-epoch brick's $(x,y)$, 
forcing the chosen reference stars to have zero mean proper motion. 
The quadratic terms in the solution were meant to model systematic errors, 
either from the plate (uncorrected Optical Field Angle Distortion) or from Tycho-2 proper motion 
systematics\footnote{Such systematics 
were indeed found later on in some tests. The Tycho-2 Catalog, although an astrometric 
catalog based on late epoch space-based data, has early epoch positions from  
ground-based data that are known to suffer from significant magnitude 
equation.} unavoidably propagated backwards in time into the computed 1st-epoch $(x,y)$.
These solutions yield measured relative proper motions in each brick.
The vast majority of the solutions only needed linear terms
and the typical standard error of the solutions varied from about 5 to 8 mas yr$^{-1}$,
which is dominated by the intrinsic proper motion dispersion of the reference stars.

The reference stars were chosen with the following criteria: 
$1<V-J<1.5$, $0.25<J-K_s<0.65$ and $13<K_s<15$, which according to 
\cite{2000ApJ...542..804N} isolates mostly G-M dwarfs in the Galaxy disk, 
located between 0.4 to 1.6 kpc from us, with an estimated mean distance 
of $\approx$ 650 pc. Their distribution in $V$ magnitude
ranges mostly from $V=15$ to $V=16.5$. 
In general there are between 200 and 500 reference stars per field, 
depending on the galactic latitude (our fields extend from $b\approx-50^o$ to 
$b\approx-20^o$). The intrinsic proper
motion dispersion of the Galactic disk reference stars was seen to increase
with $|b|$, reflecting the changing kinematics of the Galaxy along it.

\begin{figure}
\centering
\includegraphics[width=4in,angle=270]{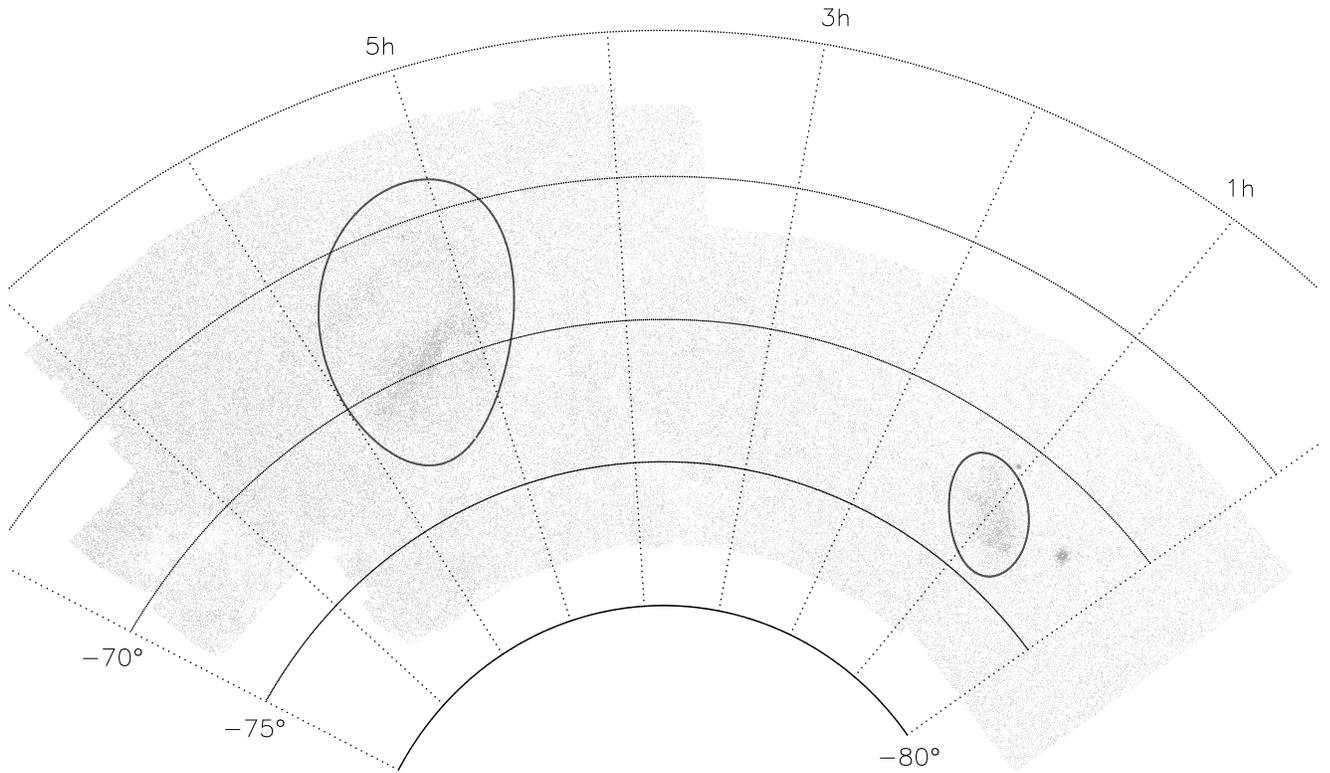}
\caption{Sky map of the relative proper-motion reference stars. 
The grey  curves enclose the so called contaminated areas.}
\label{fig_refmap}
\end{figure}

Despite the photometric cuts to select the reference stars, 
contamination by LMC and SMC stars could not be avoided in the densest parts of the Clouds. 
This forced us for the time being to restrict our investigation to those fields in
which we trust the relative proper motions, as being measured with respect to
bona fide Galactic foreground stars. These areas were defined as shown
in Figure \ref{fig_refmap}. Although a substantial number of Magellanic Clouds 
stars were lost in this selection, on the other hand, confusion due to image crowding 
at these locations render these fields useless anyway, due to the 
risk of misidentifications.

From the initial 13880 bricks available, 12180 are in the non-contaminated
areas. After rejecting bricks from the discarded plates (Section \ref{ss_plateobs}),
10900 are left to build the catalog of proper motions, mostly outside the Magellanic Clouds.
In order to increase the number of LMC and SMC stars measured at the end, 
contaminated fields that overlap with this non-contaminated catalog were later on directly 
tied into it, and common stars had their proper motions averaged.

\subsection{Combining the proper motions}\label{sc_overlap}

Once relative proper motions with respect to the Galactic foreground stars
had been measured in the non-contaminated area, different approaches were tried to combine them
into a single global well-defined reference frame. Reference stars
were chosen hoping that their mean motion along the sky could be
described by a smooth function. This goal was indeed attained as confirmed
by the fact that the measured relative proper motion of all known extragalactic objects
were very precisely fit by a quadratic polynomial in $(\alpha,\delta)$.

At this point, applying this polynomial to the relative proper motions
to convert them to absolute, and then averaging all measurements, is
a way to combine all the information available per star.
But this would yield a catalog with a rather noisy 
zero point as one moves along the sky, mostly due 
to the real intrinsic proper motion dispersion of the reference stars and the fact
that two overlapping frames may not have the same reference stars.
Given two frames with about 50\% overlap between them, means
that both have 50\% of the reference stars in common, while
the other 50\% are different. Given two samples with $N$ data points
each, both with the same dispersion $\sigma$ and both 
having 50\% of their points in common, it can be easily shown that
their individual mean values typically differ by $\sqrt{\frac{2}{N}}\sigma$.
In other words, we can expect that two overlapping frames
typically differ in their relative proper motion systems by $\approx 0.6$
mas yr$^{-1}$, for $N=220$ and $\sigma=6.5$ mas yr$^{-1}$.
This is too large for the level of precision that we want to obtain.

A variant of the so-called Block Adjustment solutions
\citep{1960AN....285..233E,1979AJ.....84.1775J,1981RMxAA...6..115S,
1988ApJ...334..465E,1988AJ.....96..409T} was considered to link
every frame's relative system into a global one.
Each frame, containing a sample of the whole population of reference
stars, realizes a local system that could deviate from the global system,
due to statistical, measurement and/or systematic errors. A linear function
per frame was considered sufficient to describe the difference between the frame
and the global system. Therefore three pa\-ra\-me\-ters per frame, per proper-motion component
($\mu_\alpha\cos\delta$ and $\mu_\delta$ are solved separately) need to be determined.
The coefficients of the quadratic polynomial that globally describes the mean motion
of the reference stars are determined as well. To 
simultaneously solve several thousands frames, 
we would need to invert an approximately  $40,000\times40,000$ matrix 
to get the parameters values and their errors. 
For this project, such a scheme was deemed impractical at this time.
Nonetheless, ideas about using this approach over smaller areas first,
like the SPM fields, and then performing another block adjustment solution
to join these regions, are being adopted for future work.

A more practical approach was considered to bring each individual frame's
reference system {\it closer} to the global one. 
A single frame's reference system is statistically more deviant 
from the global system than is the system defined
by all stars, since the proper-motion reference stars are only a subset of all stars.
Using all stars in common between overlapping frames, we can adjust
each frame's proper motions to agree with the average of the surrounding fields. 
This adjustment also helps to correct residual distortions, as they
are statistically smoothed out in the average frame.
To avoid frames drifting away from the global system as they are being 
aligned to one another, all reference stars in the
field are explicitly assigned a relative proper motion of zero.
Once the adjustments are applied, new averages can be computed, and the
whole process is iterated until the adjustments converge to zero.

\begin{figure}
\centering
\includegraphics[height=5.5in,angle=270]{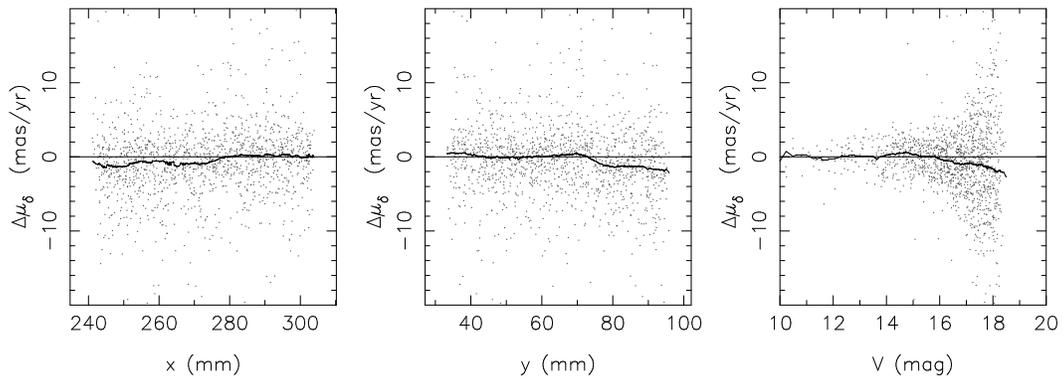}
\caption{Systematics when comparing a single frame with the average frame. 
This example illustrates the case of a PV frame, for which
$\Delta\mu_\delta=\mu_{\delta,\mbox{single frame}}-\mu_{\delta,\mbox{average frame}}$ 
exhibits visible trends, as shown by the median line.}
\label{fig_intoave}
\end{figure}

Before making these adjustments, we first checked to see if systematic 
differences existed as a function of the magnitude.
A non-negligible number of frames exhibited systematic trends with magnitude when compared to 
the average frame (See Figure \ref{fig_intoave}). In general, a linear function of the coordinates in the
field would take care of the geometrical distortion, but the magnitude equation required a smoothed-localized 
median, which can trace the general trend better than any parametrized fit.
Therefore, each frame is first corrected for its differential magnitude
equation with respect to the average frame, and then we proceed to
correct for the distortion, following the iterative procedure
explained above. Each frame has typically about 1500 to 2000 stars, with some
of them having up to 5000 stars, to compute the adjustments. 

With these improved relative proper motions, the quadratic
polynomial that describes the mean reflex proper motion of the extragalactic
objects was computed and used to transform the relative proper motions
into absolute ones. Polynomials for $\mu_\alpha\cos\delta$ and 
$\mu_\delta$ as functions of $\alpha$ and $\delta$ were calculated separately,
using proper-motion-error-weighted least squares.
A total of 5351 external galaxies were used across the 450 sq-degree area and
the formal errors of the polynomials, computed at the center of LMC
and SMC, based on the full covariance matrix, amount to 0.03 mas yr$^{-1}$ 
and 0.06 mas yr$^{-1}$ respectively.

At this stage, a final absolute proper motion and its corresponding error are obtained 
per star, from the error-weighted average of all the individual absolute proper motions
obtained for it. A total of 1337050 objects in the non-contaminated fields had 
final proper motions, including a good number of LMC and SMC stars located in the 
outskirts of these galaxies. For such reason, it is named the {\it outside catalog}.

In order to increase the number of LMC and SMC stars with measured absolute
proper motions, contaminated fields that overlapped with the outside catalog just
obtained, were directly tied into it by computing a linear solution
to correct their relative proper motions and put them into its absolute system.
A total of 678 additional fields were added with this procedure, eventually increasing 
the number of LMC and SMC stars by about 30\% and 50\%, respectively. 

\subsection{Zero point global correction of the Absolute Proper Motions}\label{sc_zerop}

Since we are now theoretically on the absolute reference frame defined by the external galaxies, 
our catalog should be -within measurement errors- in the same reference frame system 
of other known catalogs of absolute proper motion. When we checked 
our measures $(\mu_\alpha\cos\delta,\mu_\delta)$ of 1356 Hipparcos stars
against the Hipparcos Catalogue, we found a significant
difference between the two, that amounts to
\begin{eqnarray}
\mu_{\alpha\cos\delta,Hipparcos}-\mu_{\alpha\cos\delta,This\; work} &=& -0.49 \pm 0.07 \;\;\mbox{mas yr}^{-1} \label{hipa_err}\\
\mu_{\delta,Hipparcos}-\mu_{\delta,This\; work} &=& -1.21 \pm 0.07 \;\;\mbox{mas yr}^{-1} \label{hipd_err}
\end{eqnarray}

The source of this systematic difference could be indeed in any (or both) of 
the two catalogs. Although Hipparcos is the most accurate optical
astrometric catalog published so far, it has significant correlations between the astrometric
parameters (position, proper motion and parallax) of different stars,
when they are less than about 5 degrees apart on the sky, 
and also between the astrometric parameters for a given
star, due to the special measurement principle of Hipparcos \citep{1997ESASP1200.....P}.
In fact, a new reduction by \cite{2007ASSL..350.....V} was performed
to correct some systematic correlations in the data. However no significant difference 
was found between the old reference frame (used in this investigation) 
and the new one. 

On the other hand, our plate measurements of the galaxies are not error-free.
Tests were run to compare the final absolute proper motions if only galaxies 
with 2nd-epoch CCD data, or all of them, were used to compute the quadratic 
polynomials to transform relative proper motions into absolute. 
The polynomial for $\mu_\alpha\cos\delta$ exhibited noticeable differences 
around the SMC. This was not completely unexpected given that at that location, 
four SPM fields have plate-only data, but in any case it points to the
fact that galaxies in the 2nd-epoch plates may introduce problems. 
Hence, galaxies in the 1st-epoch plates may do the same.

As determined by \cite{1998AJ....115..855G}, the SPM plate material exhibits
magnitude equation for galaxies and for stars that differ in functional form.
However, the two could be brought into approximate agreement by adding an offset
of -0.7 to the magnitudes of galaxy images before calculating the magnitude-equation
correction as determined from stellar images. In the present study, we have
applied the same -0.7 mag offset to galaxy images for the purpose of magnitude-equation
correction only.

A comparison of the $V$ magnitudes for the galaxies in our work, 
showed that plate photometry returned signficantly fainter magnitudes than in
CCD calibrations. Not suprisingly, this may have affected the magnitude equation
correction in the plates, since it was the calibrated plate photographic magnitude
that was used for such purpose. More surprising though, was to find that 
Eq. \ref{hipa_err} showed a linear trend versus $\alpha$ with a slope
of about 15 $\mu$as yr$^{-1}$ per degree (a bit smaller when using only
galaxies with CCD data).
The LMC and SMC center of mass positions are separated by about $20^o$,
for which the trend above indicates a zero point shift of 0.3 mas yr$^{-1}$,
that could in principle be related to Hipparcos' systematics.
On the other hand, systematics in Eq. \ref{hipd_err}
versus $\delta$ were also seen, that look to be related to the plates location
and layout.

The investigation of the SPM plates magnitude equation done
by \cite{1998AJ....115..855G} found that magnitude equation terms\footnote{A polynomial
in $(X,Y,m)$ is used to describe the magnitude equation correction.} varied
more or less uniformly from 1st to 2nd-epoch plates, so we can expect 
more or less uniform offsets with Hipparcos proper motions, if any residual uncorrected 
magnitude equation were still present in the data. 
Consequently, Eqs. \ref{hipa_err} and \ref{hipd_err} were applied to all the absolute proper motions, putting
our catalog on the system of the ICRS via Hipparcos. The inaccuracy of its zero point is dominated mostly by
Hipparcos' systematic error of 0.25 mas yr$^{-1}$, since the quadratic
polynomial, as defined by the external galaxies, was in general very accurate,
being within 0.1 mas yr$^{-1}$ of error for most of the 450 square-degree area studied.

\subsection{Final Catalogue of Proper Motions - Evaluation of Errors}
\begin{figure}
\includegraphics[width=4in,angle=270]{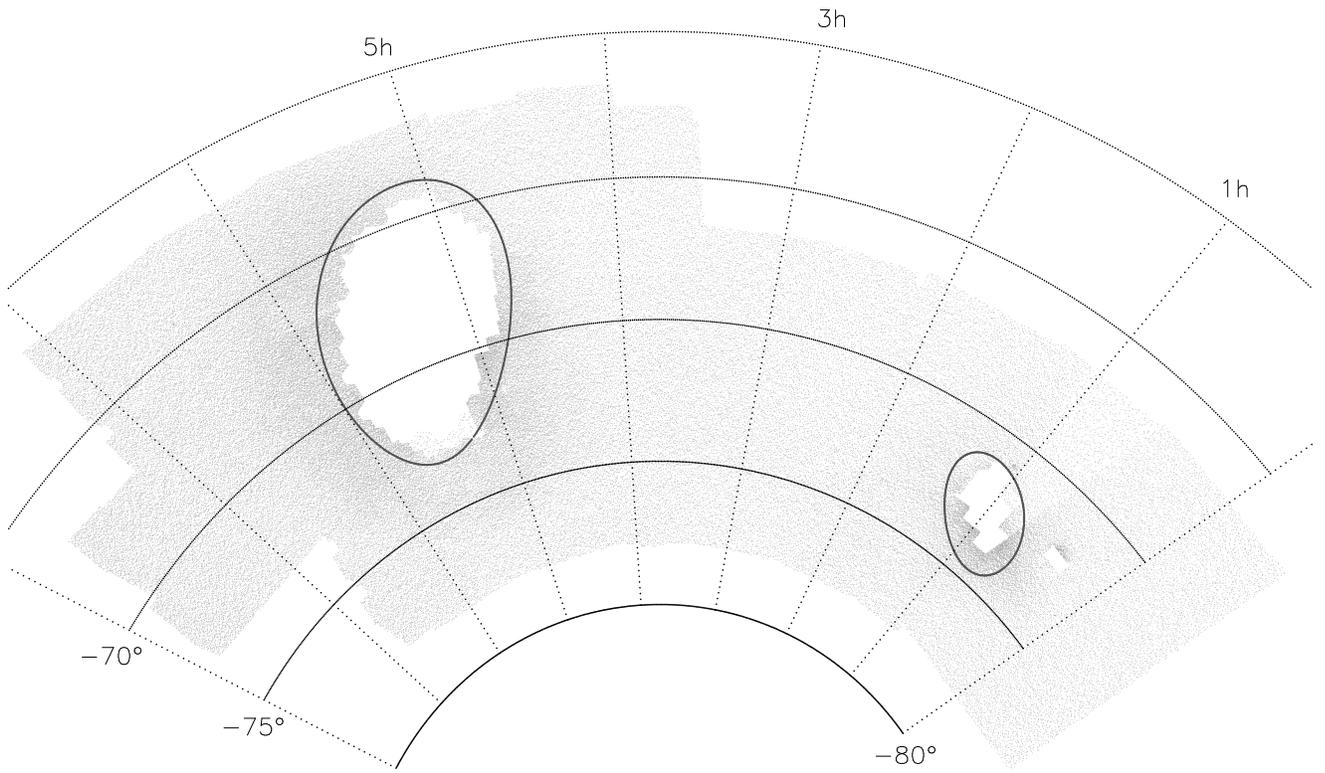}
\caption{Sky coverage of the final catalog of proper motions.
Every tenth entry is plotted in light gray. 
The ``contaminated'' areas are as in Figure \ref{fig_refmap}. The lower density area north of the SMC 
corresponds to an area where several poor quality plates were discarded.}
\label{fig_catmap}
\end{figure}

The final catalog of absolute proper motions at this point has
1,448,438 objects, with the following data listed: $\alpha, \delta, V$;
$V-J, J-K_s, H-K_s$ when available; absolute proper motions $\mu_\alpha\cos{\delta}, \mu_\delta$ 
and their formal errors $\epsilon_{\mu_\alpha\cos{\delta}},\epsilon_{\mu_\delta}$ 
in mas yr$^{-1}$, number of data points used, number of data points rejected 
(outliers were rejected based on their normalized errors), a flag to
indicate Hipparcos, Tycho-2, 2MASS extended sources, confirmed LEDA Galaxies
and QSOs, a flag to indicate if the object is or is not a reference star,
and the 2mass ID when available.
The overall distribution of the stars in the catalog can be seen in Figure \ref{fig_catmap}.

In the final catalog of proper motions, stars with $V<12$ have 
formal proper motion errors of about 0.5 mas yr$^{-1}$, and well measured stars with $12<V<15.5$, have
values that range from 0.5 to 1.3 mas yr$^{-1}$. These only reflect measurement errors, 
as they are based on the positional errors and the epoch difference.
When proper motions for the same star measured in different 
bricks are averaged together, statistical and 
systematic deviations between the bricks must be added to get the true 
error. The iterative method applied in Section \ref{sc_overlap} was designed
to reduce those deviations, but cannot make them zero. 

\begin{figure}
\includegraphics[width=4in,angle=270]{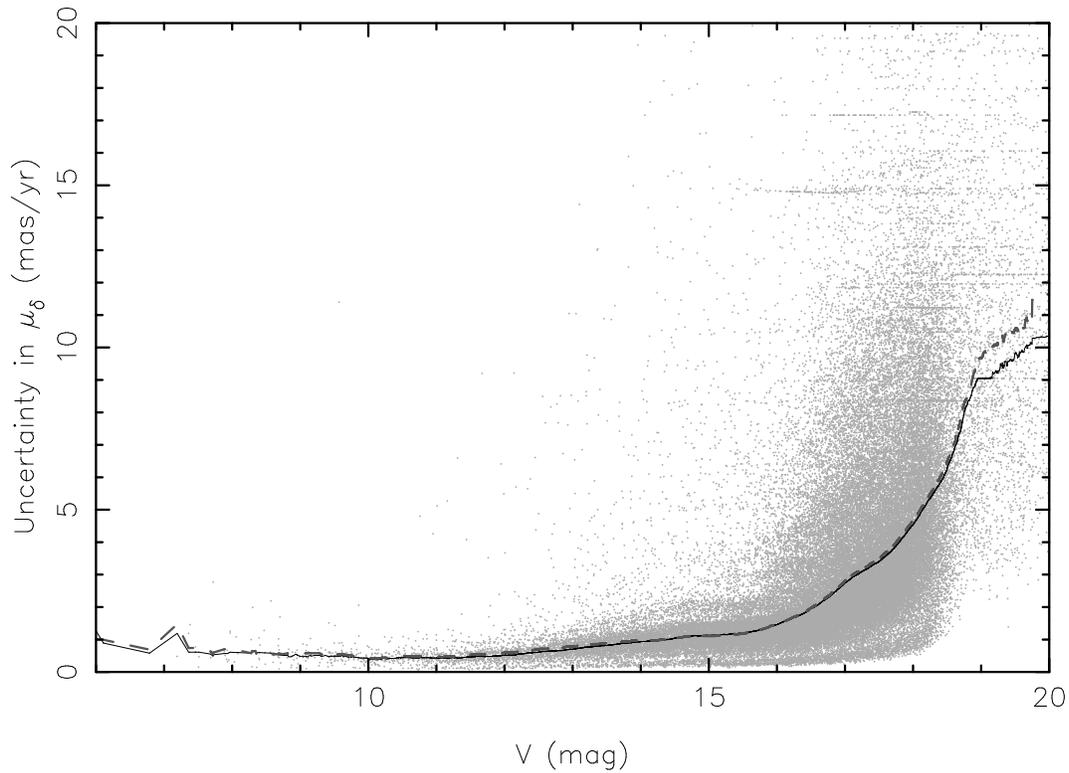}
\caption{Uncertainty in $\mu_\delta$ vs.~magnitude. 
The error estimate is based on the scatter of the individual data points
that were error-weighted-averaged into the final proper motion. The black line
is a moving median, computed over a 0.5 mag interval. The gray dashed line
is the moving median value for $\mu_\alpha\cos{\delta}$. This plot is based on a random sample 
of 5\% the size of the whole catalog.}
\label{fig_caterrors}
\end{figure}

A better way to determine the real proper motion uncertainties, is to compute 
the standard deviation of the final error-weighted average proper motions. 
Figure \ref{fig_caterrors} shows these scatter-based proper motion errors,
for a random sample of about 5\% the size of the whole catalog. 
These errors can still be an underestimate of the real proper
motion uncertainty, because in the error-weighted average the data points 
are not independent, two overlapping bricks reduced into the same plate
can produce (positively) correlated proper motions.
In this case, the true uncertainty is larger than the measured one.

The most reliable assessment of the proper motion errors  
is obtained from a comparison with external catalogs.
The scatter observed in the differences between our proper motions and those
from Hipparcos, is the combined result of both catalogs' proper motion errors.
Given that Hipparcos errors are about 1 mas yr$^{-1}$,
the measured dispersion indicates that our real proper motion uncertainties are about 2.3 mas yr$^{-1}$, for
stars brighter than $V=10$. This coincides well with the scatter-based errors 
in Figure \ref{fig_caterrors}, at the bright end.

The scatter in the proper motion of the LMC and SMC samples can
be used to estimate the proper motion uncertainties, at the mean magnitude
of the Clouds, since their intrinsic internal velocity dispersion\footnote{30 
km s$^{-1}$ (0.13 mas yr$^{-1}$ at 50 kpc distance) or less in each \citep{1997CAS....29.....W}.}
makes a very minor contribution to the observed scatter. Results indicate that
our real proper motion uncertainties are about 3.8 mas yr$^{-1}$
for stars around $V=16.4$, entirely consistent with the scatter-based
uncertainties in Figure \ref{fig_caterrors}.

At the faint end, the dispersion in the proper motion of external galaxies
varies substantially, depending on whether they have CCD or plate 2nd-epoch data. 
In the first case, the dispersion is 11 mas yr$^{-1}$
at a mean magnitude of $V=17.3$. In the second case, the dispersion is 21 mas yr$^{-1}$ 
at a mean magnitude of $V=18.5$. Since the plate photometry for the galaxies produced 
systematically fainter magnitudes than the CCD photometry, the difference in 
these two sets reflects also the difference in precision between (better) CCD and
(worse) plate measurements, and not just the increase of errors with magnitude.
At these magnitudes, our scatter-based proper motion
uncertainties are about 50\% below these values.

In general, the scatter-based proper motion errors are a rather good indicator
of the real uncertainties in the proper motions for stars brighter than
about V=16.5. More importantly,
all the above external estimates: 2.5, 3.8 and 11 mas yr$^{-1}$ 
at $V\approx 10, 16.4$ and 17.3, respectively,
are smaller in size than what was achieved in SPM3, clearly
showing the increased precision due to having 2nd-epoch CCD data. 

\section{Proper Motion of the Magellanic Clouds}\label{sc_pmmcs}

\subsection{Selection of LMC and SMC dominated samples}\label{sc_pmmcs_sel}

\begin{figure}
\centering
\plottwo{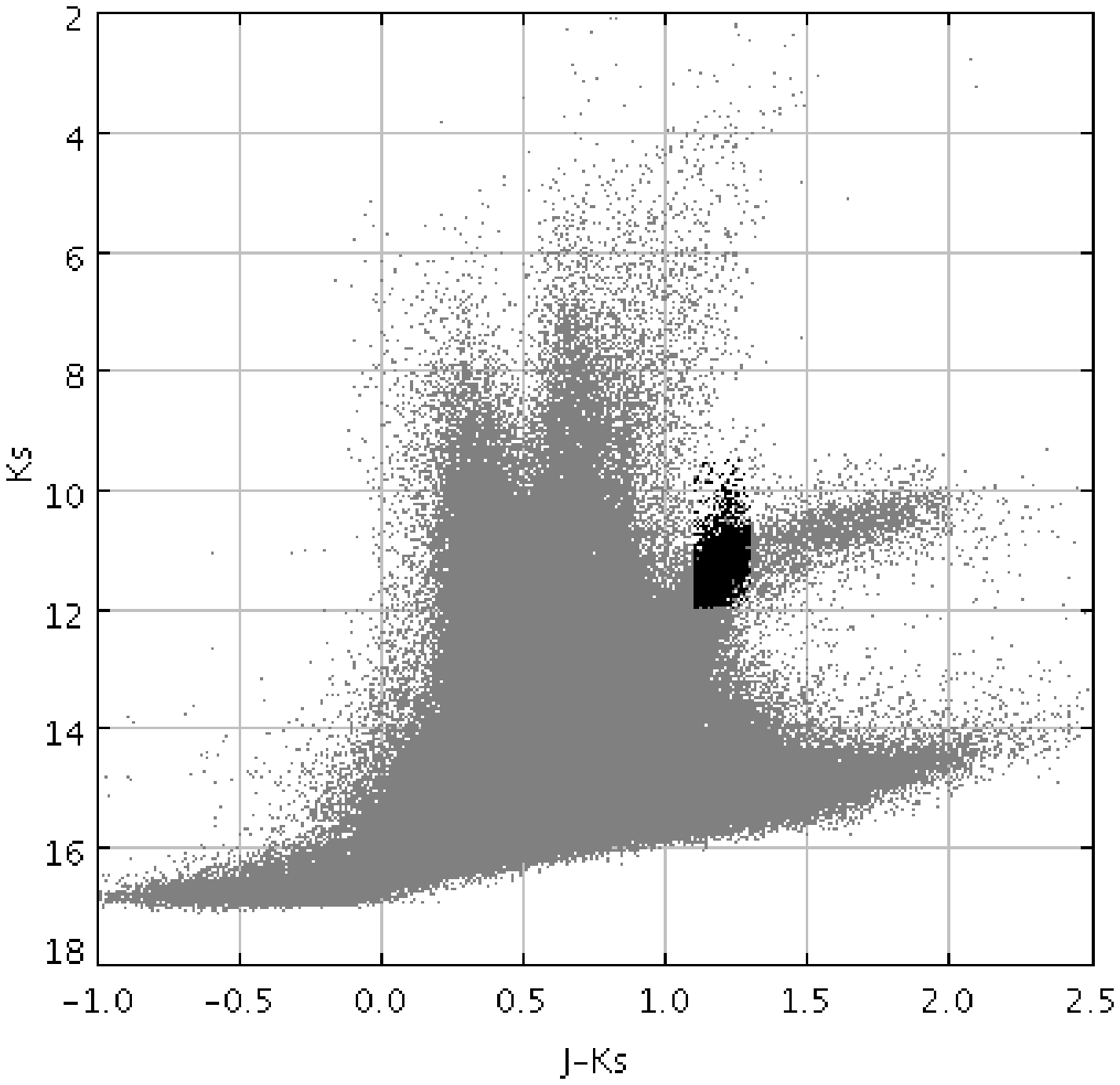}{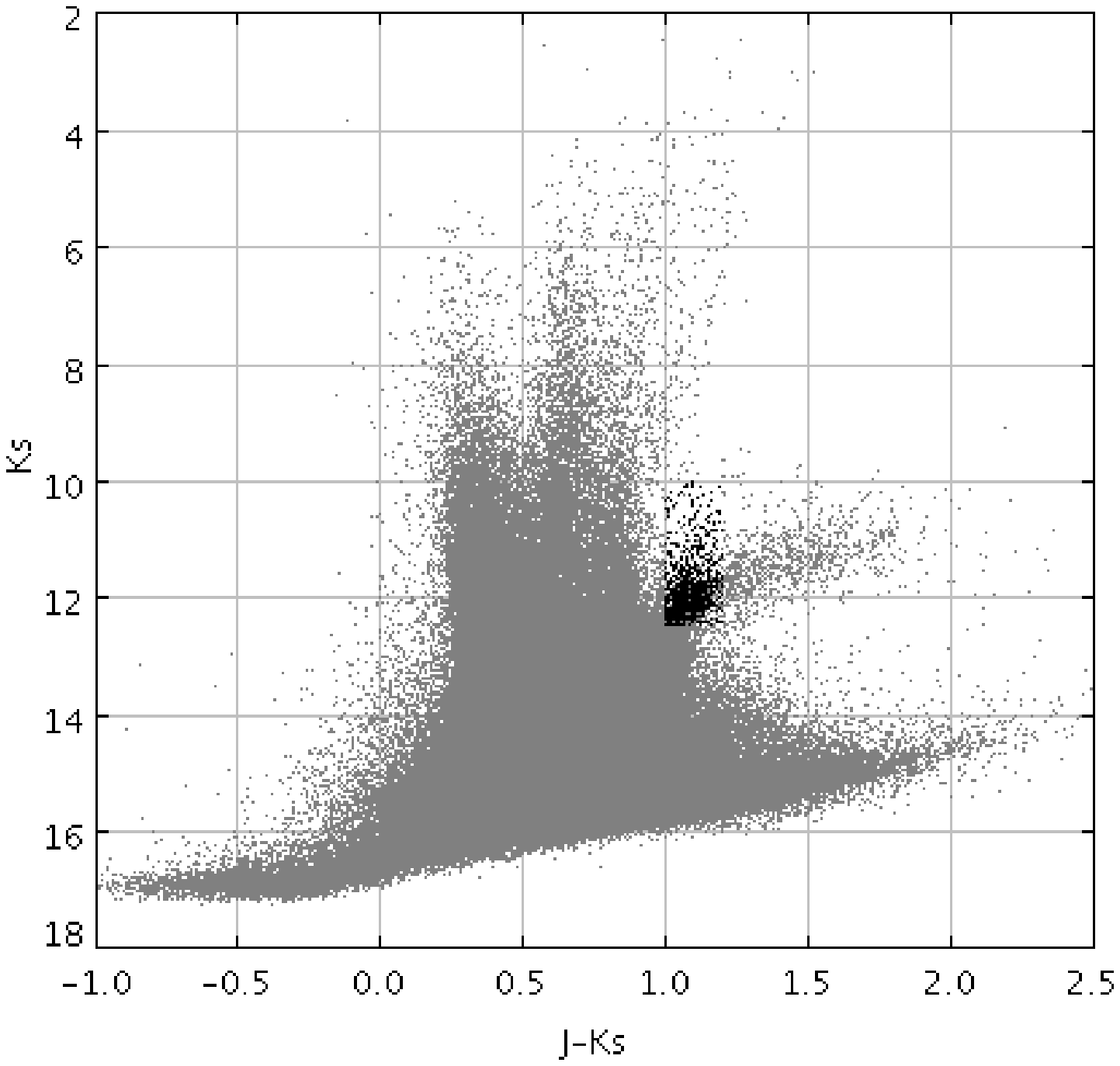}
\caption{Infrared CMDs in the vicinity of the LMC (left) and SMC (right) 
based on 2MASS photometry of the stars 
in our proper motion catalog. The two bluest vertical sequences visible
in both CMDs are dominated by foreground Galactic stars, 
while the bright red sequences are dominated by LMC/SMC stars. 
Black points on each planel show the red giant star
samples selected to measure the mean motion of each Cloud.}
\label{fig_cmdclouds}
\end{figure}

A photometric selection was made to choose bona fide red giant LMC and 
SMC stars, based on the analysis of the 2MASS LMC 
infrared color magnitude diagram (CMD) of \cite{2000ApJ...542..804N} (their sample ``J''). 
The photometric cuts applied, as seen in Figure \ref{fig_cmdclouds}, are:
\[
\begin{array}{lcccc}
\mbox{LMC :} &  1.1\leq J-K_s \leq 1.3 & \mbox{ and } & 9.5\leq K_s\leq 12 \\
\mbox{SMC :} &  1.0\leq J-K_s \leq 1.2 & \mbox{ and } &  10\leq K_s\leq 12.5
\end{array}
\]
Only stars with CCD 2nd-epoch data were selected, as stars whose proper motions 
were based on plate data only showed a significanly higher dispersion and some visible
systematics. We were also forced to discard an area of high stellar density,
at $71^o\leq\alpha\leq 76^o$ and $-68^o\leq\delta\leq -71^o$,
with CCD data close to the LMC center, which consistently showed deviant results 
probably caused by misidentifications. 

\begin{figure}
\centering
\plottwo{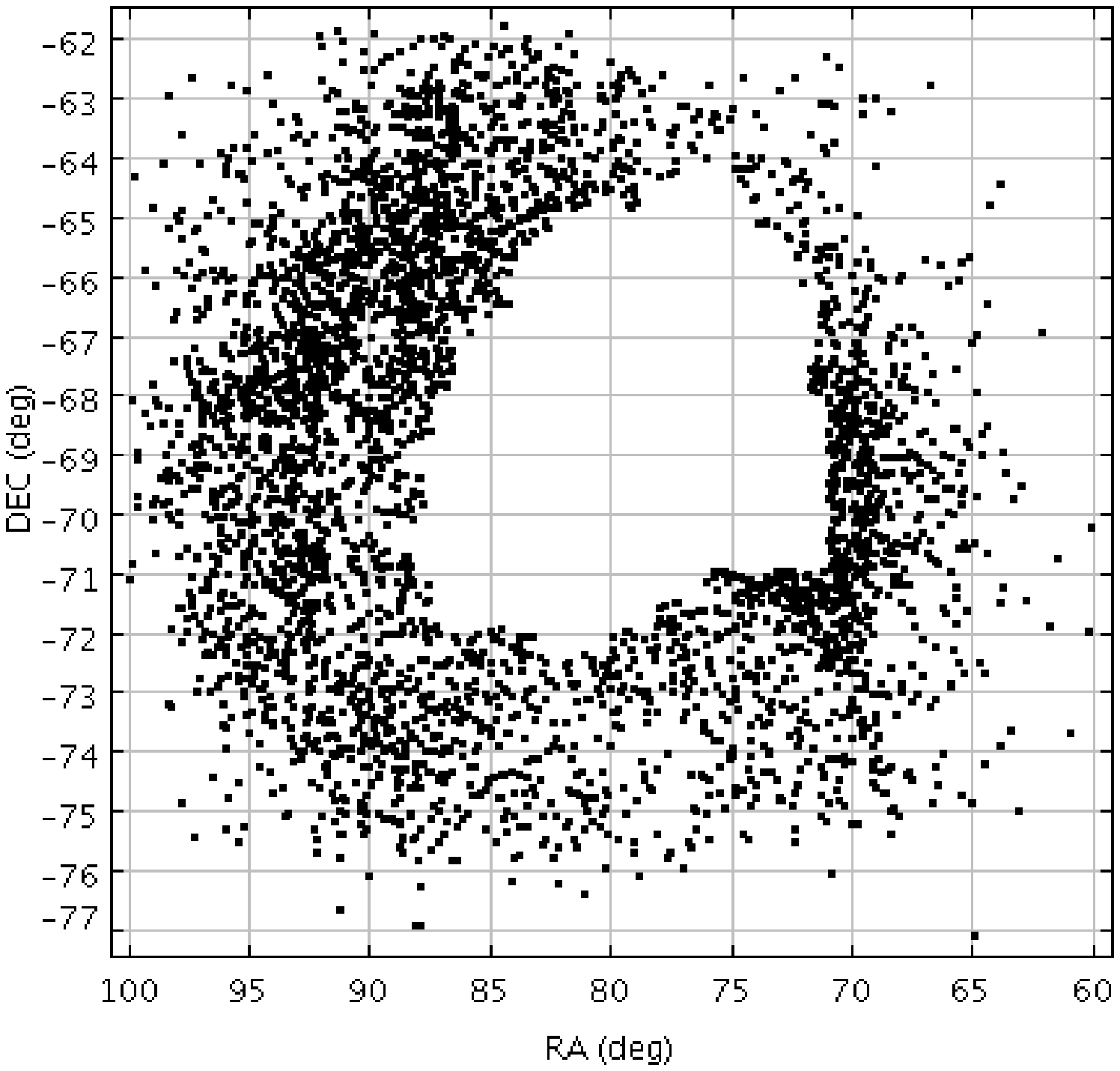}{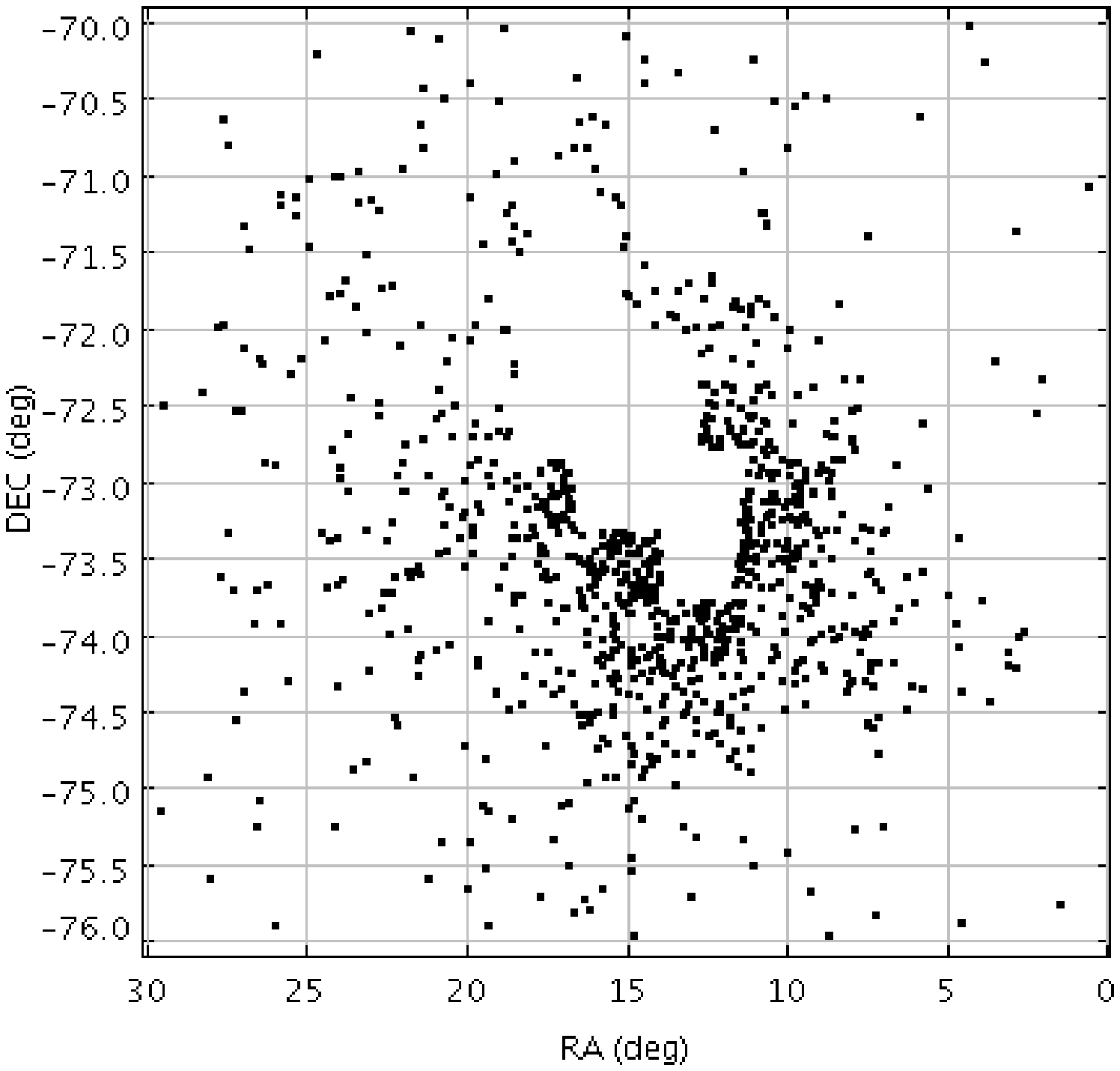}
\caption{Distribution on the sky of LMC and SMC sample stars 
selected from our catalog of proper motions.}
\label{fig_posclouds}
\end{figure}

\begin{figure}
\centering
\includegraphics[width=\textwidth]{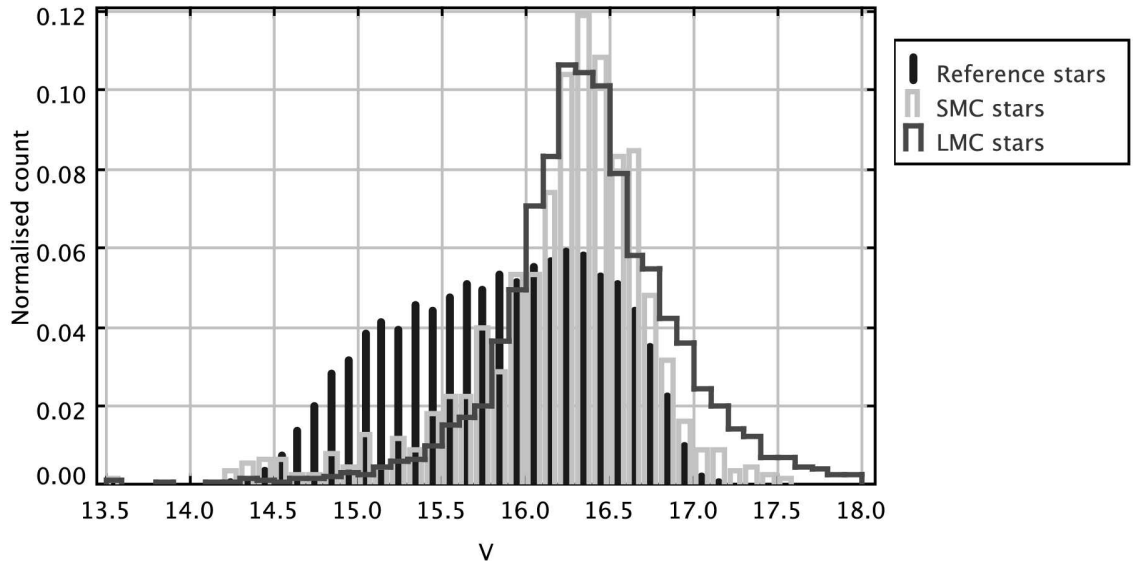}
\caption{Histogram of $V$ magnitude for the reference stars, LMC and SMC stars.
All three data sets overlap substantially in $V$, which helps to reduce
systematics in the proper motions related to brightness.}
\label{fig_histmag}
\end{figure}

3822 LMC and 964 SMC stars were selected, as seen in Figure \ref{fig_posclouds}, 
to measure the mean absolute proper motion of the Clouds. Bluer sequences of the CMDs 
in Figure \ref{fig_cmdclouds} containing LMC/SMC stars 
have signficant contribution from Milky Way stars and therefore were not considered.
The redder sequences of LMC/SMC
AGB stars have very faint $V\approx 17.5$ magnitudes, 
consequently the proper motion errors are too large to be useful.
Also, given the magnitude-related problems in the plates, 
it is desirable to have the smallest possible difference in brightness
between the reference stars and the Clouds' stars, and indeed, the
chosen samples overlap sufficiently in magnitude (See Figure \ref{fig_histmag}).

\subsection{Absolute proper motion of the LMC and SMC}\label{sc_pmmcs_lmc}

Probability plots \citep{hamaker1978} of the chosen samples yielded the 
mean and dispersion values for the LMC and SMC proper motion listed in Table \ref{tab_motion}.
The errors quoted include: the formal error of the mean value
($\sigma/\sqrt{N_{stars}}$), the error of the quadratic 
polynomial at the LMC and SMC centers, 
transformation to Hipparcos errors (Eqs. \ref{hipa_err} and \ref{hipd_err}), 
and the estimated Hipparcos systematic error (0.25 mas yr$^{-1}$). 
As explained before, the error budget is dominated by Hipparcos systematics.

\begin{figure}
\centering
\includegraphics[scale=0.7]{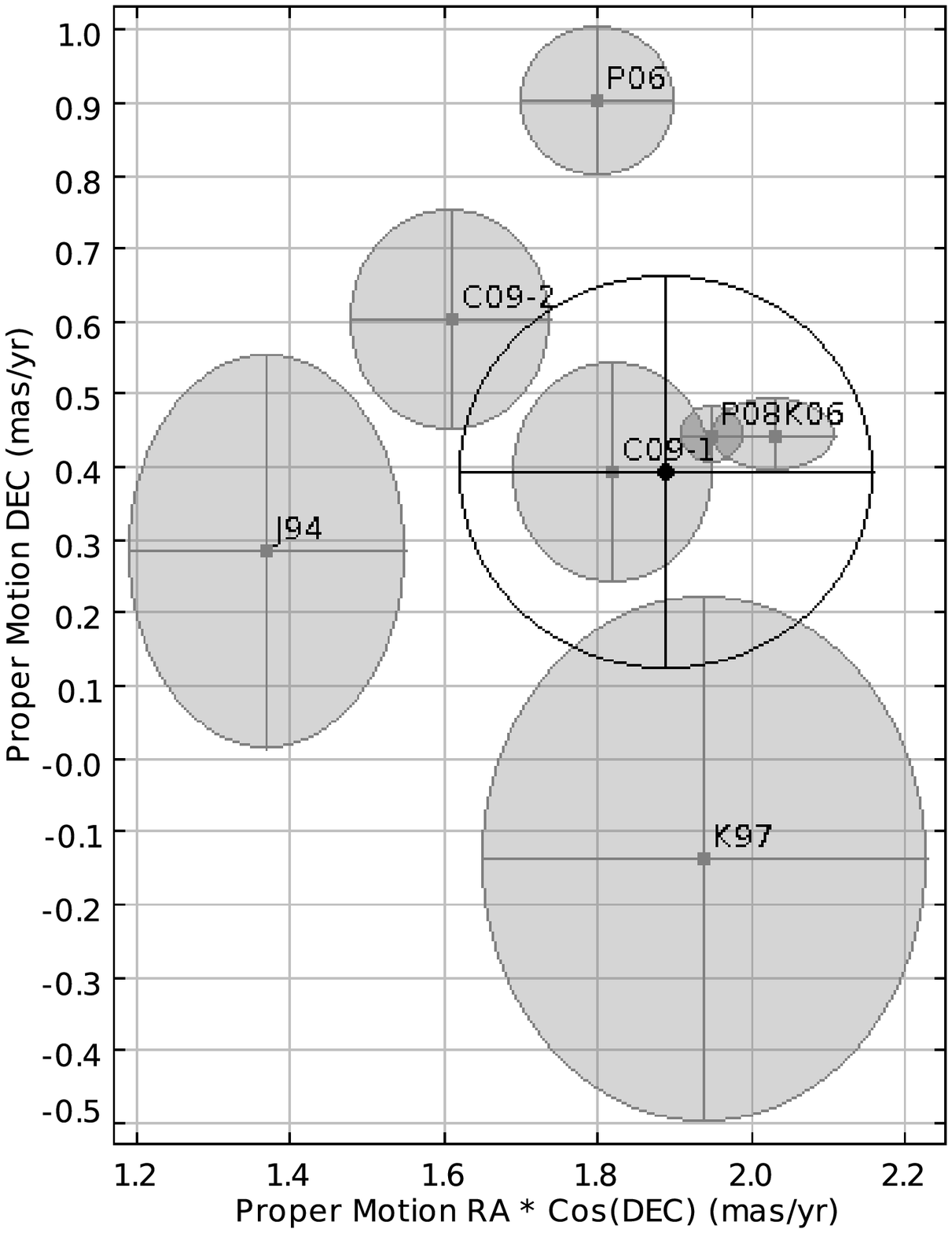}
\caption{Proper Motion of the Large Magellanic Cloud. Labels are as follows:
J94=\cite{1994AJ....107.1333J}, K97=\cite{1997ESASP.402..615K},
K06=\cite{2006ApJ...638..772K}, P06=\cite{2006AJ....131.1461P},
P08=\cite{2008AJ....135.1024P}, C09-1=\cite{2009AJ....137.4339C} (1), 
C09-2=\cite{2009AJ....137.4339C} (2).
All these results are plotted in gray while ours is plotted in black.}
\label{fig_lmc_compare}
\end{figure}

\begin{figure}
\centering
\includegraphics[width=\textwidth]{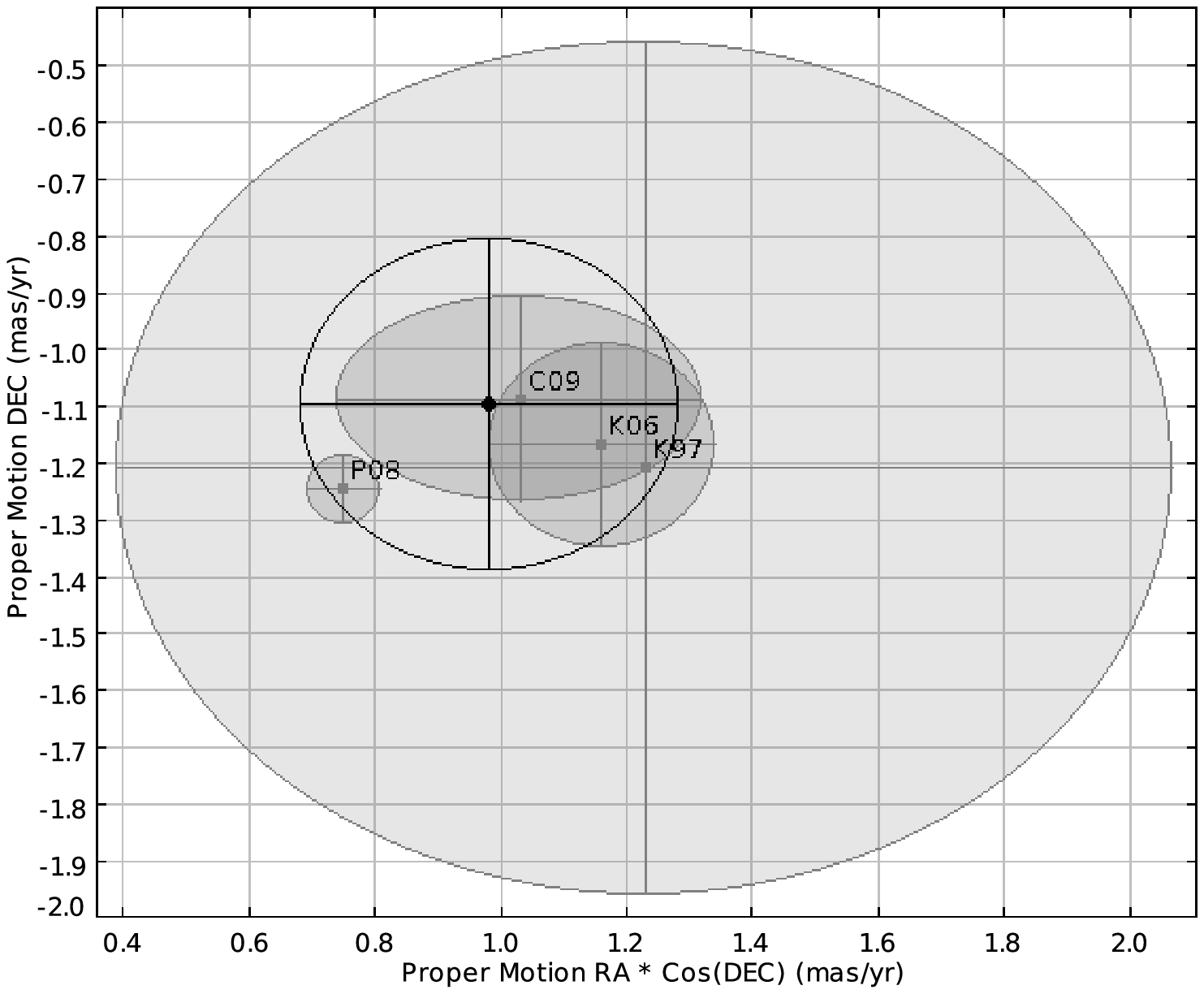}
\caption{Proper Motion of the Small Magellanic Cloud. Labels are as follows:
K97=\cite{1997ESASP.402..615K}, K06=\cite{2006ApJ...652.1213K},
P08=\cite{2008AJ....135.1024P},C09=\cite{2009AJ....137.4339C}.
All these results are plotted in gray while ours is plotted in black.}
\label{fig_smc_compare}
\end{figure}

Table \ref{tab_compare} and Figures \ref{fig_lmc_compare} and \ref{fig_smc_compare} 
summarize how our results compare with recent measurements of the proper motion of 
the Magellanic Clouds. The error bars hinder a more precise conclusion 
about the individual tangential velocities of the Clouds based on our data. 
Nonetheless, the methodology used to measure the stellar proper motions
in our catalog permits us to make a rather precise measurement of the 
proper motion of the SMC with respect to the LMC, as explained in more detail
in Section \ref{ss_cloudsrelpm}.

\subsubsection{Center-of-mass proper motion}\label{sc_rotdis}

The large extent of the Clouds and their non-negligible depth means that 
all previous investigations, which measured proper motions on scattered small fields,
had to convert their measured values into a  {\it center-of-mass proper motion}.
That is because a given space velocity at a fixed distance projects 
differently on radial velocity and proper motion at different locations in the sky, 
following the same principle of the Moving Cluster method and the Solar 
Motion\footnote{One of the first attempts to estimate the
proper motion of the Magellanic Clouds was done by measuring gradients
in the radial velocity along them \citep{1977A&A....57..265F,1988ApJ...327..651M}}.
Besides, proper motion obviously scales with distance.

Additionally, internal rotation must also be taken into account for the LMC.
In the case of the SMC, such correction is deeemed unnecessary because
its stellar component is mostly supported by velocity dispersion \citep{2006AJ....131.2514H}.
The LMC's rotation curve is obtained from radial velocities of Carbon stars
\citep{2002AJ....124.2639V,2007ApJ...656L..61O} and yield widely accepted values
of $V_{rot,LMC}=50-60$ km s$^{-1}$. Nonetheless,
\cite{2008AJ....135.1024P} estimates its own  $V_{rot,LMC}=120$ km s$^{-1}$, 
based on the gradient of their measured proper motions along the radius in the LMC disk,
and use such value for the rotation correction.

Given such discrepant values, \cite{2009AJ....137.4339C} actually obtains 
two final results for the LMC, \cite{2009AJ....137.4339C}-(1) refers to their final proper motion 
when using $V_{rot,LMC}=50$ km s$^{-1}$, while
\cite{2009AJ....137.4339C}-(2) does so for $V_{rot,LMC}=120$ km s$^{-1}$.
For the latter, it must be noted that $\mu_\alpha\cos\delta$ deviates 
noticeably from the other determinations in Table \ref{tab_compare}.

Altogether, the typical correction for perspective
effect for the LMC from both methods is about $\pm$ 0.2 mas yr$^{-1}$ and may rise
to about $\pm$ 0.5 mas yr$^{-1}$ for locations farther than $5^o$ from the LMC
center, running more or less in opposite directions at opposite locations on the Cloud.
Rotation effects for the LMC are usually less than 0.1 mas yr$^{-1}$.
Perspective effects for SMC are much smaller. For previous studies, these
corrections are necessary and in some cases yield quite different values
for the same fields, but in our work, given the spatial extent and symmetry of the data, the net effect 
on the mean motion of the Clouds is very close to zero, and no correction is done.

\subsection{Relative proper motion of the SMC with respect to the LMC}\label{ss_cloudsrelpm}

As explained in Section \ref{sc_overlap}, we measured the mean motion of our
reference stars precisely all over our field of view,
in particular at the location of LMC and SMC 
within 0.03 and 0.06 mas yr$^{-1}$,
respectively. Combined with the relative proper motion
of LMC and SMC stars with respect to these reference stars, we can
indeed measure the proper motion of the SMC with respect to that of
the LMC, with a higher precision, limited by the
error just quoted plus the formal error of the mean coming from
the number of stars and their measured scatter.

From Table \ref{tab_motion} and taking the errors quoted above into account, 
it is straightforward to obtain the relative proper motion of the SMC 
with respect to the LMC
\begin{eqnarray}
\Delta\mu_{\alpha\cos\delta}(SMC-LMC) & = & -0.91 \pm 0.16\;\; \mbox{mas yr}^{-1} \label{rel_mua}\\
\Delta\mu_{\delta}(SMC-LMC) & = & -1.49 \pm 0.15\;\; \mbox{mas yr}^{-1} \label{rel_mud}
\end{eqnarray}
These values cannot be transformed directly into a measurement 
of the relative velocity between the Clouds.
Being at different locations in the sky means that the
planes of their tangential velocities are different as well, and the
necessary rotation and projections to measure the SMC velocity on the
LMC re\-fe\-ren\-ce frame does not allow us to obtain the relative space velocity 
as merely a function of the relative proper motion between the Clouds.

\begin{figure}
\centering
\includegraphics[width=\textwidth]{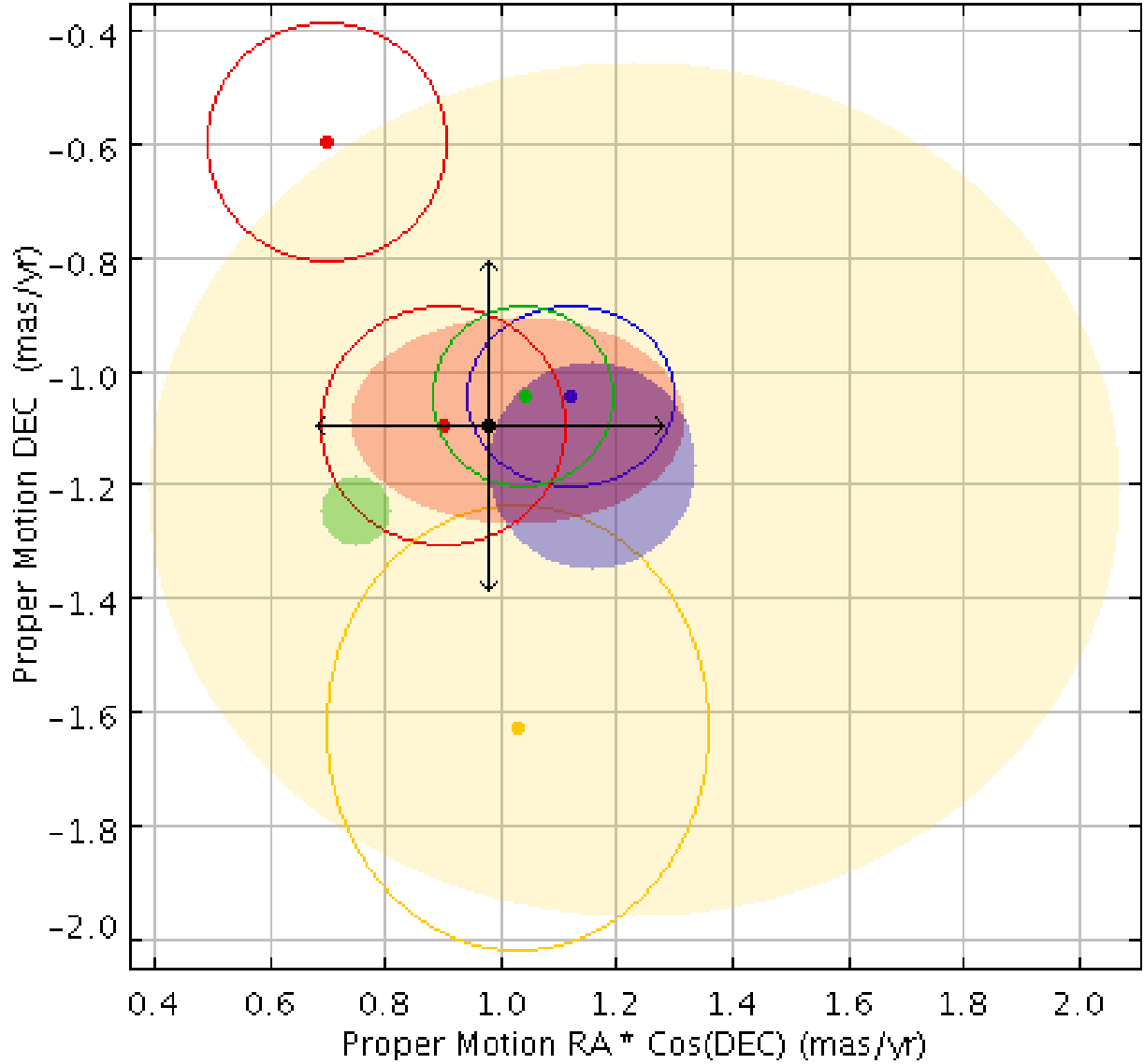}
\caption{Comparison of previous and new determinations of the Proper Motion of the SMC.
Previous determinations from Figure \ref{fig_smc_compare} 
are plotted with filled translucent symbols:
K97 in yellow, K06 in blue, P08 in green, C09 in red.
Our new determinations based on the same studies' LMC proper motion
plus our relative SMC-LMC proper motion, are plotted with open symbols 
of the same colors (C09 has two determinations of the LMC proper motion). 
Previous results are consistent with ours when the
filled and the open symbol of the same color overlap. Our SMC proper
motion in plotted with the black symbol.}
\label{fig_smc_improv}
\end{figure}

But, we can use these values to obtain new independent measurements of the SMC's
absolute proper motion, based on existing measurements of the LMC's absolute proper motion
plus our precise relative proper motion from above.
Moreover, since all authors that directly measured the proper motion of the SMC
had previously measured the LMC's proper motion as well, we can
verify if their original SMC results are consistent with our relative measure.
Figure \ref{fig_smc_improv} shows the absolute proper-motion determination for the SMC
from Table \ref{tab_smcimprov} (made by combining our relative LMC-SMC motion with
absolute LMC motions from the literature), compared to the direct determinations 
of the absolute proper motion of the SMC from Table \ref{tab_compare}.

Except for \cite{2008AJ....135.1024P} and \cite{2009AJ....137.4339C}-(2), 
all published measurements of the SMC proper motions are consistent with our new ones.
These two works are the only ones that use $V_{rot,LMC}=120$ km s$^{-1}$. 
It is worth noting that field L11 in the LMC from \cite{2008AJ....135.1024P} is also 
in \cite{2009AJ....137.4339C}, but their measured proper motions are significantly
different, beyond the quoted errors. All this makes us suspect that the rather
small quoted proper motion errors in \cite{2008AJ....135.1024P} underestimate 
their real uncertainties.

In summary, although our proper motions for the LMC and the SMC separately 
are in agreement - within error bars - with 
\cite{2006ApJ...638..772K,2006ApJ...652.1213K}, \cite{2008AJ....135.1024P} and
\cite{2009AJ....137.4339C}, our relative
proper motion of SMC with respect to LMC is consistent only
with \cite{2006ApJ...652.1213K,2006ApJ...638..772K} and \cite{2009AJ....137.4339C}-(1).

\subsection{The space motion of the clouds}\label{sc_pmmcs_spa}

The individual as well as relative space velocities of the Clouds,
as derived from our LMC and SMC proper motions, 
are given in Table \ref{tab_vel_param}. The escape 
velocity at the distance of the LMC is estimated to be
300-350 km s$^{-1}$, depending on the Galactic potential model
used (either a simple isothermal sphere or a more elaborate ``cosmologically inspired'' 
Navarro-Frank-White (NFW) dark matter profile \citep{1996MNRAS.278..191G,2002ApJ...573..597K,2008ApJ...684.1143X}.
Taken at face value, the galactocentric velocities in Table \ref{tab_vel_param} 
indicate that the LMC is traveling at a speed that is very close to the
escape velocity, while the SMC is still below the escape velocity of the Galaxy. 
Unfortunately, the uncertainties hinder a more definitive conclusion regarding their binding status.

From our proper motion measures, we determine that the SMC is moving at 89 $\pm$ 54 km s$^{-1}$ 
with respect to LMC. Our error bars do not
allow us to determine whether or not the Clouds are bound to each other, given that the escape
velocity from the LMC at the SMC location is about 90 km s$^{-1}$ (assuming a simple point-mass geometry, 
a mass for the LMC of  $2\times 10^{10} M_\odot$ and 23 kpc for the distance between the Clouds).
Yet, we can use our new SMC proper motions to obtain additional
determinations of the relative velocity between the Clouds, using the more precise
LMC proper motions available in the literature, and the other needed input parameters
listed in Table \ref{tab_vel_param}. Then we obtain more estimates of $||(U,V,W)_{SMC-LMC}||$,
which are listed in Table \ref{tab_relvel}.
For comparison, this table also lists the relative velocity originally quoted 
by the references used.

\begin{figure}
\centering
\includegraphics[width=\textwidth]{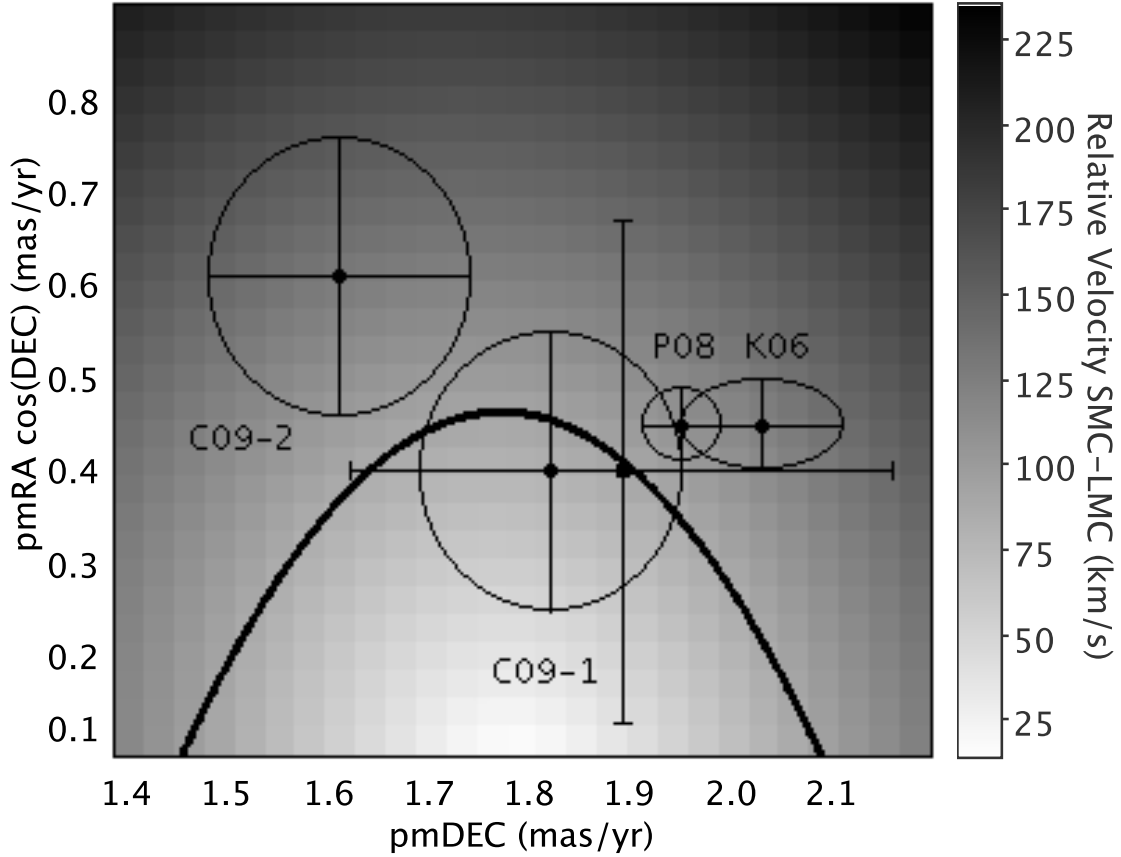}
\caption{Relative velocity between the Clouds as a function of the absolute proper motion
of the LMC. For any given value $(\mu_\alpha\cos\delta,\mu_\delta)_{LMC}$ in the diagram, 
$(\mu_\alpha\cos\delta,\mu_\delta)_{SMC}$ is obtained from Eqs.
\ref{rel_mua} and \ref{rel_mud} and then $||(U,V,W)_{SMC-LMC}||$ is computed 
(other needed parameters from Table \ref{tab_vel_param})
and plotted with a grey color indicative of its value. LMC proper motions and errors from
K06, P08, C09-1 and C09-2 are plotted in black with their error ellipses, and from this work 
with error bars only. The black parabollic curve indicates the values of LMC proper motion for which
$||(U,V,W)_{SMC-LMC}||=90$ km s$^{-1}$. Values of LMC proper motion that
fall below this curve roughly correspond to the Clouds forming a 
bound binary system.}
\label{fig_grid_dv}
\end{figure}

It is important to be aware of how sensitive $||(U,V,W)_{SMC-LMC}||$ is for different values of 
the LMC proper motion (See Figure \ref{fig_grid_dv}). It is 
clear that even the smallest proper motion errors 
quoted translate into a substantial relative velocity error between the Clouds. 
On top of that, the measured values are close enough to the escape velocity
of the SMC with respect to the LMC, such that any conclusion regarding the binarity of
the Clouds is still far from decided. The same situation applies to the individual
space velocities of the Clouds, the binding status of the
Clouds to the Milky Way is also extremely sensitive to the individual proper motions.

Thus, although our individual absolute proper motions for the Clouds cannot
tell us much about their orbits, our measured relative proper motion between 
LMC and SMC allowed us to identify the best self-consistent measurements of the proper motion 
of both clouds, from those authors that had measured both galaxies with 
claimed very good accuracies. Our measured space velocity has also helped
to check which results were more consistent with that value, 
since it is quite sensitive to the proper motions used.

\section{Implications of our results to the dynamics of the Magellanic System}\label{sc_implications}

The most prevalent scenario for the orbit of the Magellanic Clouds, before the
HST proper motion results, favored them as a binary system in a bound
orbit around the Milky Way. Dynamical simulations were performed 
under the presumption that the Clouds were bound to each other,
due to the existence of the common HI envelope surrounding them \citep{1963AuJPh..16..570H}.
Crude timing estimates suggested that for its creation and survival, a long time of 
shared orbits was needed. The extent of the Magellanic Stream implied that the Clouds 
had undergone multiple orbits gravitationally bound to the Milky Way and to each other.

The orbits of the Clouds are naturally a key ingredient in such dynamical modeling.
Once an orbit is chosen, then a full simulation of the Clouds themselves is run.
The main goal has been to reproduce as much as possible the position and 
radial velocity of the Magellanic Stream, 
though significant validation is obtained if the Leading Arm, 
the Magellanic Bridge, and the distorted structure of the Clouds, 
are also replicated. Another ``condition'' imposed is that the 
Clouds share the same orbital plane with the Magellanic Stream,
since the latter runs more or less on a great circle on the sky.

Depending on the particular questions under study, Cloud model investigators
use different approaches. Either one or both Clouds are considered, 
using collisionless (stars) and/or collisional (gas) particles.
The Cloud(s) is(are) made of one or many particles,
which can be massless test particles under a given potential, or
a conglomerate of self-gravitating particles or even
``sticky'' self-gravitating particles (to model hydrodynamic
processes). 

Overall, two competing scenarios have been systematically studied
and have had some degree of success in the modeling of the mentioned 
structures: the tidal model and the ram-pressure model. In the first scenario,
the Galaxy extracted a tidal plume from the LMC and/or the SMC, which gave
origin to the Magellanic Stream, in a previous close encounter of the three bodies
about 1.5 Gyr ago. A most recent encounter 200 Myr ago created the 
Magellanic Bridge. The second scenario proposes that the Stream and the Leading
Arm consist of material that has been ram-pressure-stripped from the LMC 
(and SMC), during its last passage through the extended ionized Halo of the Galaxy,
about 500 Myrs ago. 

A complementary scenario to extract substantial amounts of gas and no stars from the Clouds,
the blowout model, has been proposed by \cite{2008ApJ...679..432N}. The main hypothesis 
is that star formation outflow in the leading edge of the LMC has been blowing out or puffing up the gas over 
the past 2 Gyr, making it easier for ram pressure and tidal forces to strip it off.

The first dynamical simulation of the Magellanic System 
\citep{1980PASJ...32..581M} was done knowing only the radial
velocity of the Clouds, and very little about the mass of the Milky Way
at the Clouds' distances. Some dynamically permissible parameters\footnote{
By specifying the orbital inclination of the LMC with respect to the Galactic
Plane and its perigalactic distance, the equations of energy and angular momentum
conservation may be solved to yield the three components of the LMC space
velocity.} were assumed in order to
have all the input parameters necessary to compute an orbit. 
Their models required the inclusion of a massive halo (an idea that was just starting to
being accepted then) in order to reproduce the highly negative radial velocities observed
in the tip of the Magellanic Stream.  Conversely, assuming that the LMC was in a bound orbit 
provided estimates consistent with a massive Galactic halo. 

In general, all the simulations for the Magellanic Clouds
\citep{1982MNRAS.198..707L,1994MNRAS.266..567G,1994MNRAS.270..209M,1994A&A...291..743H,
1995ApJ...439..652L,1996MNRAS.278..191G,2003MNRAS.339.1135Y,2005MNRAS.363..509M,
2005MNRAS.356..680B,2006MNRAS.371..108C,2007ApJ...668..949B}
are based on the backwards integration of the equations of motion, first
applied by \cite{1980PASJ...32..581M}. Most papers consider an isothermal
sphere with a given constant rotational velocity at a large galactocentric
distance, while three of the most recent calculations use NFW dark matter halo profiles.
Not surprisingly, the mass (profile and amount) of the Galaxy is an important
source of uncertainty in the orbital models. The masses of the Clouds 
are as well another source of error, their current distorted state makes
any dynamical estimate of the mass a difficult task.

Given the large mass of the Clouds, dynamical friction\footnote{The ``retardation'' of a moving object
when it passes through a region with non-vanishing mass density,
caused by its gravitational interaction with the particles of that
continuous mass \citep{1943ApJ....97..255C}.} 
can significantly reduce their perigalacticon distance as they move through the
Galactic Halo. N-body simulations of self-gravitating particles
naturally consider it by default, while in other cases, it is accounted for by
using an analytical expression given by \cite{1987gady.book.....B}.
Recent studies \citep{2003ApJ...582..196H,2005A&A...431..861J}
have found though, that the latter tends to circularize orbits to
excess when compared to equivalent N-body simulations, thus
some simulations have scaled down its effects. 

The important point to consider here is that a high orbital eccentricity,
or equivalently a high transverse motion with respect to the Galaxy, 
is needed to lead to the formation of the high-velocity Magellanic Stream. 
Therefore, the initial conditions of the 
Clouds' orbits must be such that even after the effects of dynamical friction, orbital
eccentricity is high enough for the Stream to be formed.
Interestingly enough, \cite{1995ApJ...439..652L} had pointed out that
a hyperbolic encounter of the Clouds with the Galaxy, in which
they are passing by for the first (and only) time, could lead to the
tidal stripping of gas segments that would later infall and trail
rapidly behind. Such a model was discarded then, on the basis that the 
Clouds were presumed bound to the Galaxy. 

Another source of dynamical friction, the LMC dark matter halo,
is considered as well in the calculations of the Magellanic orbits
\citep{2005MNRAS.356..680B}. Its effects are quite important on the binarity of the galaxy pair,
exerting a significant frictional drag on the SMC when it penetrates 
the LMC halo. If they get close enough, they would merge quickly, and since 
this has not occurred yet, it implies either that this force is
negligible or that the Clouds became bound to each other relatively
recently \citep{1994MNRAS.266..567G}. \cite{2005MNRAS.356..680B}
also found that under these conditions, the Clouds cannot keep their
binary status for more than $\approx$5 Gyr in the past.
This opens the possibility of the Clouds being coupled only recently.

Curiously, the very first dynamical model
of the Magellanic System \citep{1980PASJ...32..581M} had to make an
extensive search for the binary state that could produce the Magellanic
Stream. The choice of tangential velocities for SMC had to be
so specific, that they could - theoretically -
predict it within $\pm$ 5 km s$^{-1}$. This prompted
them and others to make the first inferred estimates of the proper motion of the
Clouds. This early model though, did not include a dark matter halo for the LMC.

A similar situation was faced later on by \cite{1994A&A...291..743H},
who explain that if the perturbing forces acting on the LMC and/or
SMC have to be small enough to leave the binary system intact while
simultaneously producing long streams similar in shape to the Stream,
then the evolution time of these streams has to be rather long and very special,
and properly chosen initial conditions for the test particles
in the simulations have to be adopted.

One point in which all orbital models agree, is that
the Clouds had a recent encounter, sometime between 200 to 500 Myr ago.
In fact, several models also found that the binarity
of the Clouds, regardless of how long it has been in place,
was most probably broken at this last collision, which happened
very close to their perigalacticton distance. In other words,
the Milky Way's powerful gravitational tide has disrupted
the pair. 

In general, searching for the appropriate binary-bound orbits for the 
Clouds that can later be used to reproduce the Magellanic System,
has been a difficult fine tuning task. The challenge is now even
harder, since the most recent HST measurements of the LMC and
SMC proper motion \citep{2006ApJ...638..772K,2006ApJ...652.1213K,2008AJ....135.1024P}
seem to indicate that the Clouds are traveling too fast to have ever been
bound to the Milky Way. 

Numerical simulations of the LMC orbit by \cite{2007ApJ...668..949B},
based on these new numbers, suggest that the Large Cloud is ``plunging'' in a highly eccentric
parabolic orbit, on its first passage about the Milky Way.
At such speed, dynamical friction is negligible,
but the choice of the Galactic potential (isothermal sphere
vs. NFW) introduces dramatic changes in the orbital history of the LMC.
In an isothermal sphere, the LMC has indeed a bound orbit, although
with an increased period and apogalacticon distance compared to
previous models. In an NFW profile, the ``best case scenario''
of a bound orbit has a period of about a Hubble time, and reaches
an apogalacticon distance of $\approx$ 550 kpc. 

Despite such 3D differences, their projected orbits on the sky
were in good agreement, so their predicted location compared to 
the Magellanic Stream's great circle, could not be used to distinguish
Milky Way mass profile models. More importantly though, in both
cases the projected orbit did not trace the Magellanic Stream,
deviating from it by about 10$^o$ on the sky.
Adding the SMC into the calculations did not reduce the disagreement,
nor using the weighted average of all pre-HST proper motions.
\cite{2007ApJ...668..949B} argues then that the usual criteria to validate 
a dynamical model of the Magellanic System, its ability to match an 
orbit with the Stream, is no longer acceptable.

The conclusion of \cite{2007ApJ...668..949B}, that the Clouds are on their 
{\it first} passage relies heavily on the large value for $\mu_\alpha\cos\delta$ that was measured
with HST for the LMC. \cite{1995ApJ...439..652L} and others have in fact argued 
that it is hard to explain how a bound LMC could have mantained
a high angular momentum perpendicular to the Galaxy's rotation axis for so long.
\cite{1989MNRAS.240..195R}, \cite{1992ApJ...386..101S} and \cite{1994AJ....107.2055B} have suggested
that an early interaction with M31 could be the source of such a high
tangential velocity. In any case, such a condition is easier to understand
with an LMC not bound to the Milky Way. In addition, the Clouds are
the only gas-rich dwarf galaxies at small galactrocentric distances 
\citep{2006AJ....132.1571V}, different from the rest in the
Local Group. All these reasons are used by \cite{2007ApJ...668..949B} to support the case
of an unbound LMC.

Notably, all LMC proper motion results after \cite{2006ApJ...638..772K}, including
this work, have produced lower values of $\mu_\alpha\cos\delta$. Therefore, 
a bound orbit with a long period and a large apogalacticon
distance is still a scenario compatible with the most recent results. As for
the binarity of the Clouds, a recent period of joint orbits could be enough
to explain the common features between the Clouds, that is the HI envelope,
the Bridge and even the Stream, and can account for the different star formation
history and chemical evolution in each Cloud, that point to
a separate origin and place of birth.

To conclude, the search for a realistic orbit of the Magellanic Clouds
is far from over. The space velocities obtained in this investigation are supportive
of a scenario in which the Magellanic Clouds are possibly currently unbound from each other, 
with the LMC traveling at a velocity that is high enough to make 
it nearly unbound to the Galaxy. But having nowadays a much more precise measurement
of the proper motion of the Clouds, has not facilitated our understanding
of their dynamics and has instead opened new questions and placed 
all the constructed models in doubt. 

\section{Conclusions and Future Work}\label{sc_end}

A catalog of absolute proper motions containing 1,448,438 objects has been
obtained from SPM material, supplemented with UCAC2 data. The catalog covers 
an estimated area of 450 square degrees except for the inner regions of LMC, SMC
and 47 Tucanae, where the high stellar density made it impossible to obtain accurate
cross-identification of the stars.

Samples of 3822 LMC stars and 964 stars were selected from the catalog to measure
the mean proper motion of the Magellanic Clouds. The results obtained are:
\[
(\mu_\alpha\cos\delta,\mu_\delta)_{LMC}=(+1.89,+0.39)\pm (0.27,0.27)\;\;\mbox{mas yr}^{-1}
\]
\[
(\mu_\alpha\cos\delta,\mu_\delta)_{SMC}=(+0.98,-1.01)\pm (0.30,0.29)\;\;\mbox{mas yr}^{-1}
\]

Our much more precise relative proper motions with respect to the photometrically
selected Galactic Disk dwarf stars, enabled us to obtain the proper motion of the SMC with
respect to the LMC, with significantly smaller uncertainties:
\[
(\mu_{\alpha\cos\delta},\mu_\delta)_{SMC-LMC} = (-0.91,-1.49) \pm (0.16,0.15)\;\;\mbox{mas yr}^{-1}
\]
This was used to obtain new independent and more precise proper motions
for the SMC, based on the more accurate LMC proper motions of other authors. 
It was also used to confirm if their separate measurements of the SMC 
proper motion were consistent or not with our results. 

After a comparison in an absolute and relative sense with previous proper motion results,
followed by a discussion of the orbital models of the Magellanic Clouds based on those results, 
we conclude that our proper motions are compatible with the LMC and SMC being born and formed
as separate entities, which later joined in a temporary binary state for
the past few Gigayears, being recently disrupted by the Milky Way in their
most recent perigalaticon passage about 200 Myr ago. The Clouds orbits are
{\it marginally} bound to the Milky Way, possibly following a very elongated
but still periodic orbit around the Galaxy.

The search for a realistic orbit of the Magellanic Clouds
is far from over. Having (formally) very accurate and precise space-based 
proper motions for the Clouds, has not facilitated our understanding
of their dynamics but has, instead, opened new questions and has placed
all dynamic scenarios of the Magellanic System in doubt. As of today,
it is still unclear if the Magellanic Stream and the Leading Arm are caused
mostly by a tidal interaction or are the result of the ram-pressure of
the Galactic Halo on the gas of the Clouds. 

Given the inherent difficulties in
measuring an accurate proper motion for the Magellanic Clouds,
the obvious dangers that systematic errors pose in those
measurements and the fact that the dynamical models of the Magellanic System
are extremely sensitive to small variations in the proper motions
of the Clouds, we believe that we are not yet in the position
of considering them known parameters in the orbital calculation.
But we are getting closer.

\subsection{Future work}

From the very begining, this investigation of the proper motion of 
the Magellanic Clouds was known to be restricted by the conditions 
and characteristics of the SPM material.
We have proven that even under those constraints, these data are able
to produce independent significant results on the proper motion of nearby
dwarf galaxies. Thus, there is certainly room for improvement. 

Second epoch CCD data over the whole field are needed,
especially in the inner areas of the Clouds. Since the definition of an adequate uncontaminated relative
reference frame in those areas is difficult, it is also necessary to devise
an adequate proper motion reduction method, to precisely link these 
fields into the general global relative reference frame. A scaled-down
version of the block adjustment method could be put in place here, in which
plate-size fields of view are assembled first, and then another solution is
run to paste those into the larger global system.

Since the dominant limiting factor in terms of precision is the plate measurement
errors, additional improvement can be achieved by re-scanning the first epoch plates 
used in this work with the Yale PDS, which yields 2 times
more precise positions than the currently measured ones. As
PDS scanning of a full plate is very time-consuming, a subset of 
properly pre-selected objects should be measured. 

Although additional HST follow up observations are already planned to improve
the space-based proper motions, and other research teams are still working on
additional CCD ground-based measurements, these studies are still limited since
they must correct their observed proper motions into a center-of-mass value.
Therefore, it is still worth trying to improve our results, since they offer a
wide field extended coverage of the Magellanic Clouds.

Another future work being considered, is to search in the Intercloud region
for coherent structures in the relative proper motion space, 
to identify stars whose motion is directed towards the LMC. This will
require to find what possibly is a small number of stars spread over 
an extended area that share systematic proper motions.

\acknowledgements

Vieira would like to thank Dr. Burt Jones for being the
external reader of her PhD thesis, on which this manuscript is based.

We would like to thank the many former and present members of the Cesco Observatory,
Universidad Nacional de San Juan and Yale Southern Observatory, who have
contributed to the SPM program over the years.

Vieira would also like to thank the National Science Foundation 
(grants AST04-07292, AST04-07293, AST09-08996) and the Yale 
Astronomy Dept for their financial support during her graduate career.
She and the other authors are grateful to the NSF and to Yale University
for support of the SPM and this research in particular throughout the many
years required to complete the SPM.

This publication makes use of data products from the Two Micron All Sky Survey, which
is a joint project of the University of Massachusetts and the Infrared Processing and
Analysis Center/California Institute of Technology, funded by the National Aeronautics
and Space Administration and the National Science Foundation.

\bibliographystyle{aa} 
\bibliography{ms}                        


\newpage


\begin{deluxetable}{cccccc}
\tabletypesize{\scriptsize}
\tablecolumns{6}
\tablewidth{0in}
\tablecaption{SPM plates used in this paper\label{tab_plateslist}}
\tablehead{ & & \multicolumn{4}{c}{SPM plates} \\ 
\colhead{SPM field} & \colhead{$(\alpha,\delta)_{1950}$} & 
\multicolumn{2}{c}{1st epoch} & \multicolumn{2}{c}{2nd epoch} \\ 
 & & \colhead{Plate number} & \colhead{Date} & \colhead{Plate number} & \colhead{Date}}
\startdata
 028 & $(00^h00^m,-75^o)$ & 0829B/Y & 03/NOV/72.84 & 1354/BY & 01/NOV/94.84 \\
 029 & $(01^h00^m,-75^o)$ & 0834B/Y & 06/NOV/72.85 & & \\ 
 030 & $(02^h00^m,-75^o)$ & 0826B/Y & 01/NOV/72.84 & & \\
 031 & $(03^h00^m,-75^o)$ & 0830B/Y & 03/NOV/72.84 & & \\
 032 & $(04^h00^m,-75^o)$ & 0950B/Y & 26/OCT/73.82 & & \\
 033 & $(05^h00^m,-75^o)$ & 0851B/Y & 06/JAN/73.02 & 1357/BY & 30/NOV/94.92 \\
     &                    &         &              & 1373/BY & 20/OCT/95.80 \\
 034 & $(06^h00^m,-75^o)$ & 0860B/Y & 25/JAN/73.07 &         & \\ \hline
 052 & $(00^h00^m,-70^o)$ & 0821B/Y & 06/OCT/72.77 & 1355/BY & 02/NOV/94.84 \\
 053 & $(00^h48^m,-70^o)$ & 0750B/Y & 24/AUG/71.65 & 1371/BY & 23/SEP/95.73 \\
 054 & $(01^h36^m,-70^o)$ & 0751B/Y & 25/AUG/71.65 & 1372/BY & 20/OCT/95.80 \\ 
 055 & $(02^h24^m,-70^o)$ & 0969B/Y & 26/NOV/73.90 & & \\  
 056 & $(03^h12^m,-70^o)$ & 0815B/Y & 16/SEP/72.71 & & \\ 
 057 & $(04^h00^m,-70^o)$ & 0946B/Y & 06/OCT/73.76 & & \\ 
 058 & $(04^h48^m,-70^o)$ & 0846B/Y & 01/JAN/73.00 & & \\ 
 059 & $(05^h36^m,-70^o)$ & 0854B/Y & 09/JAN/73.02 & 1350/BY & 17/NOV/93.88 \\
 060 & $(06^h24^m,-70^o)$ & 0842B/Y & 13/DEC/72.95 & & \\ 
 061 & $(07^h12^m,-70^o)$ & 0423B/Y & 20/FEB/69.14 & & \\ 
 088 & $(04^h00^m,-65^o)$ & 0824B/Y & 15/OCT/72.79 & & \\ 
 089 & $(04^h40^m,-65^o)$ & 0835B/Y & 06/NOV/72.85 & & \\ 
 090 & $(05^h20^m,-65^o)$ & 0853B/Y & 07/JAN/73.02 & 1394/BY & 16/NOV/93.88 \\
 091 & $(06^h00^m,-65^o)$ & 0781B/Y & 13/JAN/72.04 & & \\
     &                    & 0984B/Y & 27/JAN/74.07 & & \\
 092 & $(06^h40^m,-65^o)$ & 0986B/Y & 28/JAN/74.08 & & \\
\enddata
\tablecomments{Plates 0750B, 0751B, 1371B, 1357Y, and 1373Y,
were subsequently discarded, as visual examination of them revealed significant deficiencies
that would negatively impact the astrometric reduction at their locations.}
\end{deluxetable}

\newpage

\begin{deluxetable}{lrcccccc}
\tabletypesize{\small}
\tablecolumns{8}
\tablewidth{0in}
\tablecaption{Absolute Proper Motion of the Magellanic Clouds\label{tab_motion}}
\tablehead{
\colhead{Target} & 
\colhead{$N_{stars}$} & 
\colhead{$\mu_\alpha\cos\delta$} &
\colhead{$\mu_\delta$} & 
\colhead{$\epsilon_{\mu_\alpha\cos\delta}$} &
\colhead{$\epsilon_{\mu_\delta}$} &
\colhead{$\sigma_{\mu_\alpha\cos\delta}$} &
\colhead{$\sigma_{\mu_\delta}$} \\
& & \colhead{mas yr$^{-1}$} & \colhead{mas yr$^{-1}$} & \colhead{mas yr$^{-1}$} &
    \colhead{mas yr$^{-1}$} & \colhead{mas yr$^{-1}$} & \colhead{mas yr$^{-1}$} }
\startdata
LMC & 3822 & 1.89 &  0.39 & 0.27 & 0.27 & 3.76 & 3.59 \\
SMC &  964 & 0.98 & -1.01 & 0.30 & 0.29 & 4.12 & 3.82
\enddata
\tablecomments{The $\epsilon$ values are the errors of the mean $\mu$ values, 
which include: the formal errors ($\sigma/N_{stars}$),
the error of the quadratic polynomial at the LMC and
SCM centers, transformation to Hipparcos errors
(Eqs. \ref{hipa_err} and \ref{hipd_err}), and the Hipparcos systematic error
(0.25 mas yr$^{-1}$). The $\sigma$ values represent the scatter of the data
around the mean proper motion values.}
\end{deluxetable}

\newpage

\begin{deluxetable}{lccr}
\tabletypesize{\small}
\tablecolumns{6}
\tablewidth{0in}
\tablecaption{Recent Determinations of the Proper Motion of the Magellanic Clouds\label{tab_compare}}
\tablehead{\colhead{Author} & \colhead{$\mu_\alpha\cos\delta$} & \colhead{$\mu_\delta$} & \colhead{$N_{stars}$} \\
 & \colhead{mas yr$^{-1}$} & \colhead{mas yr$^{-1}$} & }
\startdata
\multicolumn{4}{c}{Large Magellanic Cloud} \\[6pt]
\cite{1994AJ....107.1333J}     & $1.37\pm 0.18$ &  $0.28\pm 0.27$ &  251 \\ 
\cite{1997ESASP.402..615K}     & $1.94\pm 0.29$ & $-0.14\pm 0.36\;\;\;$ &   33 \\ 
\cite{2006AJ....131.1461P}     & $1.80\pm 0.10$ &  $0.90\pm 0.10$ &  108 \\ 
\cite{2006ApJ...638..772K}     & $2.03\pm 0.08$ &  $0.44\pm 0.05$ &  810 \\ 
\cite{2008AJ....135.1024P}     & $1.95\pm 0.04$ &  $0.44\pm 0.04$ &  889 \\ 
\cite{2009AJ....137.4339C} (1) & $1.82\pm 0.13$ &  $0.39\pm 0.15$ &   41 \\ 
\cite{2009AJ....137.4339C} (2) & $1.61\pm 0.13$ &  $0.60\pm 0.15$ &   41 \\ 
This work                       & $1.89\pm 0.27$ &  $0.39\pm 0.27$ & 3822 \\[6pt] 
\multicolumn{4}{c}{Small Magellanic Cloud} \\[6pt]
\cite{1997ESASP.402..615K} & $1.23\pm 0.84$ & $-1.21\pm 0.75$ &   9 \\ 
\cite{2006ApJ...652.1213K} & $1.16\pm 0.18$ & $-1.17\pm 0.18$ & 177 \\ 
\cite{2008AJ....135.1024P} & $0.75\pm 0.06$ & $-1.25\pm 0.06$ & 215 \\ 
\cite{2009AJ....137.4339C} & $1.03\pm 0.29$ & $-1.09\pm 0.18$ &  44 \\ 
This work                   & $0.98\pm 0.30$ & $-1.10\pm 0.29$ & 964    
\enddata
\tablecomments{Results (1) and (2) from \cite{2009AJ....137.4339C} are 
obtained assuming two different rotational velocities for LMC, as 
explained in detail in Section \ref{ss_cloudsrelpm}.}
\end{deluxetable}

\clearpage

\begin{deluxetable}{lcc}
\tabletypesize{\small}
\tablecolumns{6}
\tablewidth{0in}
\tablecaption{New determinations of the Proper Motion of the SMC\label{tab_smcimprov}}
\tablehead{\colhead{Using the LMC} & \colhead{$\mu_\alpha\cos\delta$} & \colhead{$\mu_\delta$} \\
\colhead{proper motion of} & \colhead{mas yr$^{-1}$} & \colhead{mas yr$^{-1}$}  }
\startdata
\cite{1994AJ....107.1333J}     & $0.46\pm 0.31$ & $-1.21\pm 0.31$ \\ 
\cite{1997ESASP.402..615K}     & $1.03\pm 0.39$ & $-1.63\pm 0.39$ \\ 
\cite{2006AJ....131.1461P}     & $0.89\pm 0.18$ & $-0.59\pm 0.19$ \\ 
\cite{2006ApJ...638..772K}     & $1.12\pm 0.16$ & $-1.05\pm 0.18$ \\ 
\cite{2008AJ....135.1024P}     & $1.04\pm 0.16$ & $-1.05\pm 0.16$ \\ 
\cite{2009AJ....137.4339C} (1) & $0.90\pm 0.21$ & $-1.10\pm 0.21$ \\ 
\cite{2009AJ....137.4339C} (2) & $0.70\pm 0.21$ & $-0.60\pm 0.21$ \\ 
\enddata
\tablecomments{Results (1) and (2) from \cite{2009AJ....137.4339C} are 
obtained assuming two different rotational velocities for the LMC, as 
explained in detail in Section \ref{ss_cloudsrelpm}. Compare with
the direct determinations of the proper motion of the SMC in 
Table \ref{tab_compare}. Quoted errors include the contribution from our
measured relative proper motion.}
\end{deluxetable}

\newpage

\begin{deluxetable}{lccc}
\tabletypesize{\small}
\tablewidth{0in}
\tablecolumns{4}
\tablecaption{Space velocity of the Magellanic Clouds\label{tab_vel_param}}
\tablehead{\colhead{Parameter} & \colhead{Unit} & \colhead{LMC} & \colhead{SMC}} 
\startdata
\multicolumn{4}{c}{Measured or Adopted Quantities} \\[6pt]
$(\alpha,\delta)$ & deg &   (81.90,-69.87)  &   (13.20,-72.50)  \\
$(l,b)$             & deg & (280.526,-32.527) & (303.788,-44.628) \\
Distance          & kpc & 50.1 $\pm$ 2.3    & 62.8 $\pm$ 2.6   \\ 
Radial velocity   & km s$^{-1}$ & 262.1 $\pm$ 3.4   & 146.0 $\pm$ 0.6  \\
$(\mu_\alpha\cos\delta,\mu_\delta)$ & mas yr$^{-1}$ & (1.89,0.39) & (0.98,-1.10) \\
                                    &               & $\pm$ (0.27,0.27) & $\pm$ (0.30,0.29) \\[6pt]
$(U,V,W)_{\odot,LSR}$ & km s$^{-1}$ & \multicolumn{2}{c}{(10.00,5.25,7.17)} \\
$(U,V,W)_{LSR,gc}$ & km s$^{-1}$ & \multicolumn{2}{c}{(0,220,0)} \\[6pt]
\multicolumn{4}{c}{Derived Quantities} \\[6pt]
(X,Y,Z)  & kpc & (0.3,-41.5,-26.9) & (-16.2,-37.6,-44.1) \\
$R_{gc}$ & kpc & 41.5 & 40.9  \\
$(U,V,W)_{hc}$  & km s$^{-1}$ & (71.8,-468.1,226.1) & (51.5,-433.4,139.2) \\
         &             & $\pm$ (63.6,37.1,56.6)  & $\pm$ (82.1,70.3,62.2)  \\
$(U,V,W)_{gc}$  & km s$^{-1}$ & (71.8,-248.1,226.1) & (51.5,-213.4,139.2) \\
         &             & $\pm$ (63.6,37.1,56.6)  & $\pm$ (82.1,70.3,62.2)  \\
$||(U,V,W)_{gc}||$  & km s$^{-1}$ & $343.3 \pm 47.8$ & $259.9 \pm 68.6$ \\[6pt]
$||(U,V,W)_{SMC-LMC}||$   & km s$^{-1}$ & \multicolumn{2}{c}{89.2 $\pm$ 53.9}
\enddata
\tablecomments{$(\alpha,\delta)$ from \cite{2002AJ....124.2639V} (LMC) and \cite{2004ApJ...604..176S} (SMC).
(l,b) are transformed from $(\alpha,\delta)$. Distance from \cite{2001ApJ...553...47F} (LMC) and
\cite{2000A_A...359..601C} (SMC). Radial velocity from \cite{2002AJ....124.2639V} (LMC) and
\cite{2006AJ....131.2514H} (SMC). Solar motion $(U,V,W)_{\odot,LSR}$, 
and galactocentric velocity of the Local Standard of Rest $(U,V,W)_{LSR,gc}$, 
from \cite{1998MNRAS.298..387D}. $R_\odot=8$ kpc was adopted. 
$R_{gc}$ is the Galactocentric distance. $(U,V,W)_{hc}$
and $(U,V,W)_{gc}$ are the heliocentric and galactocentric space velocities of the Clouds.
$||(U,V,W)_{LMC,SMC}||$ is the relative velocity of SMC with respect to LMC.  }
\end{deluxetable}

\newpage

\begin{deluxetable}{lc}
\tabletypesize{\small}
\tablewidth{0in}
\tablecolumns{2}
\tablecaption{Relative velocity between LMC and SMC\label{tab_relvel}}
\tablehead{\colhead{Reference} & \colhead{$||(U,V,W)_{SMC-LMC}||$} \\
\colhead{} & \colhead{km s$^{-1}$} }
\startdata
\multicolumn{2}{c}{Originally quoted by reference} \\[6pt]
\cite{2006ApJ...652.1213K}     & 105 $\pm$ 42 \\ 
\cite{2008AJ....135.1024P}     & 142 $\pm$ 19 \\
\cite{2009AJ....137.4339C}-(1) & $\;\;$84 $\pm$ 50 \\
\cite{2009AJ....137.4339C}-(2) & $\;\;$62 $\pm$ 63 \\[6pt]
\multicolumn{2}{c}{New value using our $\mu_{SMC-LMC}$} \\[6pt]
\cite{2006ApJ...652.1213K}     & 123 $\pm$ 51 \\
\cite{2008AJ....135.1024P}     & 108 $\pm$ 51 \\
\cite{2009AJ....137.4339C}-(1) & $\;\;$83 $\pm$ 52 \\
\cite{2009AJ....137.4339C}-(2) & 152 $\pm$ 63 \\[6pt]
\multicolumn{2}{c}{This work} \\[6pt]
This work                      & $\;\;$89 $\pm$ 54
\enddata
\tablecomments{Results (1) and (2) from \citep{2009AJ....137.4339C} are
obtained assuming two different rotational velocities for LMC, as
explained in detail in Section \ref{ss_cloudsrelpm}.}
\end{deluxetable}

\end{document}